\documentclass[12pt,english,american]{article}
\usepackage[T1]{fontenc}
\usepackage[latin9]{inputenc}
\usepackage{geometry}
\usepackage{color}
\usepackage{babel}
\usepackage{amsthm}
\usepackage{amstext}
\usepackage{amssymb}
\usepackage{mathrsfs}
\usepackage{amsmath}
\usepackage{graphicx}
\usepackage{cancel}   
\usepackage[unicode=true,pdfusetitle,
 bookmarks=true,bookmarksnumbered=false,bookmarksopen=false,
 breaklinks=true,pdfborder={0 0 0},backref=false,colorlinks=true]
 {hyperref}
 \usepackage{setspace}
\hypersetup{
 linkcolor=blue,citecolor=blue, urlcolor=blue}
\makeatletter

\def\qed{$\Box$\medskip}

\newcommand{\beq}{\begin{equation}}
\newcommand{\eeq}{\end{equation}}
\newcommand{\beqa}{\begin{eqnarray}}
\newcommand{\eeqa}{\end{eqnarray}}
\newcommand{\ben}{\begin{arabicenumerate}}
\newcommand{\een}{\end{arabicenumerate}}

\def\bel{\begin{lem} } 
\def\eel{\end{lem} }
\def\bet{\begin{thm}}
\def\eet{\end{thm}}
\def\bed{\begin{defn}}
\def\eed{\end{defn} }

\def\bec{\begin{cor}}
\def\eec{\end{cor}}
\def\ber{\begin{rem}}
\def\eer{\end{rem}}

\usepackage{enumitem}		% customizable list environments
\theoremstyle{plain}
\newtheorem{thm}{\protect\theoremname}[section]
\theoremstyle{definition}
\newtheorem{defn}[thm]{\protect\definitionname}
\theoremstyle{plain}
\newtheorem{prop}[thm]{\protect\propositionname}
\theoremstyle{plain}
\theoremstyle{remark}
\newtheorem{rem}[thm]{\protect\remarkname}
\theoremstyle{plain}
\newtheorem{lem}[thm]{\protect\lemmaname}
\theoremstyle{plain}
\newtheorem{cor}[thm]{\protect\corollaryname}

\usepackage{txfonts,refstyle,xcolor}

\newref{con}{name = Conjecture\ }
\newref{prop}{name = Proposition\ }
\newref{def}{name = Definition\ }
\newref{sec}{name = Section\ }
\newref{sub}{name = Section\ }
\newref{thm}{name = Theorem\ }
\newref{lem}{name = Lemma\ }
\newref{cor}{name = Corollary\ }
\newref{fig}{name = Figure\ }
\newref{rem}{name =  Remark\ }

\usepackage{bbm}
\newcommand{\charf}{\mathbbm{1}}

\usepackage[all]{xy}

\newcommand{\xyR}[1]{%
     \makeatletter
     \xydef@\xymatrixrowsep@{#1}
     \makeatother
}

\newcommand{\xyC}[1]{%
     \makeatletter
     \xydef@\xymatrixcolsep@{#1}
     \makeatother
}

\newcommand{\ncol}[1]{\color{normalcolor}}

\makeatother

\addto\captionsamerican{\renewcommand{\corollaryname}{Corollary}}
\addto\captionsamerican{\renewcommand{\definitionname}{Definition}}
\addto\captionsamerican{\renewcommand{\lemmaname}{Lemma}}
\addto\captionsamerican{\renewcommand{\propositionname}{Proposition}}
\addto\captionsamerican{\renewcommand{\remarkname}{Remark}}
\addto\captionsamerican{\renewcommand{\theoremname}{Theorem}}
\addto\captionsenglish{\renewcommand{\corollaryname}{Corollary}}
\addto\captionsenglish{\renewcommand{\definitionname}{Definition}}
\addto\captionsenglish{\renewcommand{\lemmaname}{Lemma}}
\addto\captionsenglish{\renewcommand{\propositionname}{Proposition}}
\addto\captionsenglish{\renewcommand{\remarkname}{Remark}}
\addto\captionsenglish{\renewcommand{\theoremname}{Theorem}}
\providecommand{\corollaryname}{Corollary}
\providecommand{\definitionname}{Definition}
\providecommand{\lemmaname}{Lemma}
\providecommand{\propositionname}{Proposition}
\providecommand{\remarkname}{Remark}
\providecommand{\theoremname}{Theorem}

\addto\captionsamerican{\renewcommand{\corollaryname}{Corollary}}
\addto\captionsamerican{\renewcommand{\definitionname}{Definition}}
\addto\captionsamerican{\renewcommand{\lemmaname}{Lemma}}
\addto\captionsamerican{\renewcommand{\propositionname}{Proposition}}
\addto\captionsamerican{\renewcommand{\remarkname}{Remark}}
\addto\captionsamerican{\renewcommand{\theoremname}{Theorem}}
\addto\captionsenglish{\renewcommand{\corollaryname}{Corollary}}
\addto\captionsenglish{\renewcommand{\definitionname}{Definition}}
\addto\captionsenglish{\renewcommand{\lemmaname}{Lemma}}
\addto\captionsenglish{\renewcommand{\propositionname}{Proposition}}
\addto\captionsenglish{\renewcommand{\remarkname}{Remark}}
\addto\captionsenglish{\renewcommand{\theoremname}{Theorem}}
\providecommand{\corollaryname}{Corollary}
\providecommand{\definitionname}{Definition}
\providecommand{\lemmaname}{Lemma}
\providecommand{\propositionname}{Proposition}
\providecommand{\remarkname}{Remark}
\providecommand{\theoremname}{Theorem}

\begin{document}
\title{Lie-Schwinger block-diagonalization and gapped quantum chains: analyticity of the ground-state energy} 
  \author{S. Del Vecchio\footnote{Dipartimento di Matematica, Universit\`a di Roma ``Tor Vergata", Italy
/ email: delvecch@mat.uniroma2.it}\,,  J. Fr\"ohlich\footnote{Institut f\"ur Theoretiche Physik, ETH-Z\"urich , Switzerland / email: juerg@phys.ethz.ch}\,, A. Pizzo \footnote{Dipartimento di Matematica, Universit\`a di Roma ``Tor Vergata", Italy
/ email: pizzo@mat.uniroma2.it}\,, S. Rossi\footnote{Dipartimento di Matematica, Universit\`a di Roma ``Tor Vergata", Italy
/ email: rossis@mat.uniroma2.it}}

\date{17/08/2019}

\maketitle

\abstract{We consider quantum chains whose Hamiltonians are perturbations by interactions of short range of a Hamiltonian that does not couple the degrees of freedom located at different sites of the chain and has a strictly positive energy gap above its ground-state energy. For interactions that are form-bounded  w.r.t. the on-site Hamiltonian terms, we have proven that the spectral gap of the perturbed Hamiltonian above its ground-state energy  is bounded from below by a positive constant \textit{uniformly} in the length of the chain, for small values of a coupling constant; see \cite{DFPR}. The main result of this paper is that, under the same hypotheses,  the ground-state energy is analytic for values of the coupling constant belonging to a fixed interval, uniformly in the length of the chain. Furthermore, assuming that the interaction potentials are invariant under translations, we prove that, in the thermodynamic limit, the energy per site is analytic for values of the coupling constant in the same fixed interval. In our proof we use a new method introduced in \cite{FP}, which is based on \emph{local} Lie-Schwinger conjugations of the Hamiltonians associated with connected subsets of the chain. We prove a rather strong result concerning complex Hamiltonians corresponding to complex values of the coupling constant. }
\\
\section{Introduction: Models and Results}
In this paper, we continue our  study  of spectral properties of Hamiltonians associated with a family of quantum chains with interactions of short range. Included are bosonic systems such as an array of coupled anharmonic oscillators. For the Hamiltonians considered in this paper, we have proven that their ground-state energy is finitely degenerate and the spectral gap above the ground-state energy  is bounded from below by a positive constant, \textit{uniformly} in the length of the chain; see  \cite{DFPR}. 
The purpose of the present paper is to show that, under the hypotheses considered in \cite{DFPR}, the ground-state energy is actually analytic in the coupling constant, in a fixed disk centered at the origin that can be chosen to be independent of $N$ (number of sites in the chain).  The method\footnote{See \cite{DFFR} for the use of a similar block-diagonalization in a simpler context. Ideas somewhat similar to the scheme in \cite{FP}
have been used in work of J. Z. Imbrie, \cite{I1}, \cite{I2}.} is based on \textit{iterative conjugations} of the Hamiltonians, which serve to block-diagonalise them with respect to a fixed orthogonal projection 
and its orthogonal complement, and that carries over to the complex Hamiltonians obtained  from the physical ones by replacing the physical (real) coupling constant by a complex parameter.

Our analysis is motivated by recent widespread interest in characterising topological phases of matter;
see, e.g., \cite{MN}, \cite{NSY}, \cite{BN}.

The type of Hamiltonians considered in this paper have been studied before to prove the existence of a spectral gap above the ground-state energy (or the existence of an isolated eigenvalue for complex Hamiltonians), often using so-called ``cluster expansions'', but usually for bounded and real interactions; see \cite{DFF}, \cite{FFU}, \cite{KT}, \cite{Y}, \cite{KU}, \cite{DS} \cite{H} and refs. given there.  Regarding  the main result of this paper, i.e., analyticity of the ground-state energy, we mention that, in the paper by D. Yarotsky (see \cite{Y}) dealing with unbounded interactions, a weak*-analyticity result was proven: 
\emph{For any local observable (i.e., an operator of bounded ``lattice support"), its expectation value in the ground state of the system is analytic in the coupling constant, {\color{red}$t$}, in a disk  independent of the size of the lattice.}

\subsection{A concrete family of quantum chains}\label{intro-def}
The Hilbert space of pure state vectors of the quantum chains studied in this paper has the form
\begin{equation}\label{tensorprod}
\mathcal{H}^{(N)}:= \bigotimes_{j=1}^{N} \mathcal{H}_{j}\,,
\end{equation}
where $\mathcal{H}_{j}\simeq \mathcal{H}, \, \forall j=1,2,\dots, N$ and where $\mathcal{H}$ is a separable Hilbert space. Let $H$ be a non-negative operator with the properties that $0$ is an eigenvalue of $H$ corresponding to an eigenvector $\Omega \in \mathcal{H}$, and 
$$H \upharpoonright_{\lbrace \mathbb{C} \Omega \rbrace^{\perp}} \geq \charf \,,$$
where $\charf $ is the identity operator.

\noindent
We define
\begin{equation}\label{H_i}
H_{i}:= \charf_{1}\otimes \dots \otimes \underset{\underset{i^{th} \text{slot}}{\uparrow}}{H} \otimes \dots \charf_{N}\,.
\end{equation}
By $P_{\Omega_i}$ we denote the orthogonal projector onto the subspace
\begin{equation}\label{vacuum_i}
\mathcal{H}_{1}\otimes \dots \otimes \underset{\underset{i^{th} \text{slot}}{\uparrow}}{\lbrace \mathbb{C} \Omega \rbrace}\otimes \dots \otimes \mathcal{H}_{N} \subset \mathcal{H}^{(N)}\,, \quad \text{  and}\quad   P_{\Omega_i}^{\perp} := \charf - P_{\Omega_i}\,.
\end{equation}
Then 
\begin{equation*} 
H_i = P_{\Omega_i} H_i P_{\Omega_i} + P_{\Omega_i}^{\perp} H_i P_{\Omega_i}^{\perp} \,,
\end{equation*}
with
\begin{equation}\label{gaps}
P_{\Omega_i} H_i P_{\Omega_i}=0\,,\quad P_{\Omega_i}^{\perp} H_i P_{\Omega_i}^{\perp} \geq P_{\Omega_i}^{\perp}\,.
\end{equation}
In \cite{DFPR} we have studied quantum chains on the graph $I_{N-1,1}:= \lbrace 1, \dots, N\rbrace, \,N< \infty$ arbitrary, with Hamiltonians of the form
\begin{equation}\label{Hamiltonian}
K_{N}\equiv K_{N}(t):= \sum_{i=1}^{N} H_i + t \sum_{\underset{k \leq \bar{k}}{I_{k,i}\subset I_{N-1,1}}} V_{I_{k,i}}\,, 
\end{equation}
where $t\in \mathbb{R}$ is a coupling constant, $\bar{k} < \infty$ is an arbitrary, but fixed integer,
$I_{k,i}$ is the ``interval'' given by $\lbrace i, \dots, i+k\rbrace, \, i=1, \dots, N-k$, \, and $V_{I_{k,i}}$ is a symmetric operator acting on $\mathcal{H}^{(N)}$ with the property that 
\begin{equation}\label{potential}
V_{I_{k,i}} \,\, \text{  acts as the identity on  }\,\, \bigotimes_{j\in I_{N-1,1}\,,\, j\notin I_{k,i}} \mathcal{H}_{j}\,.
\end{equation}
We call $I_{k,i}$ the ``support'' of $V_{I_{k,i}}$. It is assumed that 
\begin{equation}\label{klmn-cond}
D((H_{I_{k,i}}^0)^{\frac{1}{2}})\subseteq D(V_{I_{k,i}}),\quad \quad |\langle \phi\,,\, V_{I_{k,i}} \phi \rangle|\leq a\langle \phi\,,\,(H_{I_{k,i}}^0+1)\phi\rangle \,,
\end{equation}
for any $\phi \in D((H_{I_{k,i}}^0)^{\frac{1}{2}})$  where $H_{I_{k,i}}^0:=\sum_{l=i}^{i+k}H_l$, for some universal constant $a>0$. Under these assumptions,  and using the inequality
\begin{equation}
\sum_{I_{k,i}\subset I_{N-1,1}}H_{I_{k,i}}^0\leq (k+1)\sum_{i=1}^{N} H_i\,,
\end{equation}
one shows that, for $|t|$ sufficiently small (depending on $\bar{k}$ and $a$, but independent of $N$), the symmetric operator in (\ref{Hamiltonian})  is defined and bounded from below on $D(H^0_{I_{N-1,1}})$. It can be extended to a densely defined self-adjoint operator whose domain we denote $D(K_N)\subseteq D((H_{I_{N-1,1}}^0)^{\frac{1}{2}})$, namely the Friedrichs extension, which is uniquely determined by the property $D(K_N)\subseteq D((H_{I_{N-1,1}}^0)^{\frac{1}{2}})\equiv Q(K_N)$, where $Q(K_N)$ is the form domain. It is not difficult to check that,  under our hypotheses on the potentials, this extension coincides with the self-adjoint operator defined through the KLMN theorem,  starting from the closed quadratic form associated with  (\ref{Hamiltonian}).
%ensures that: 1)   there is a domain that we denote  $D(K_N)$  where $K_N$ is a well defined self-adjoint operator; 2) the  form domain $Q(K_N)$ coincides with the form domain, $Q(H_{I_{N-1,1}}^0)\equiv D((H_{I_{N-1,1}}^0)^{\frac{1}{2}})$,  of $H_{I_{N-1,1}}^0$. Hence, we deduce that $D(K_N))\subseteq D((H_{I_{N-1,1}}^0)^{\frac{1}{2}})$. 

In the present paper, we will consider a complex coupling constant, $\tau$,  instead of $t\in \mathbb{R}$, and analyze the closed quadratic form given by
\begin{equation}\label{complex-Hamiltonian}
\sum_{i=1}^{N} H_i + \tau \sum_{\underset{k \leq \bar{k}}{I_{k,i}\subset I_{N-1,1}}} V_{I_{k,i}}:=\kappa_N \equiv \kappa_{N}(\tau)\,.
\end{equation}
For $|\tau|$ sufficiently small (depending on $\bar{k}$ and $a$,  but independent of $N$), we have the following general results (see Theorems 3.9 and 2.1 of \cite{K}), starting from the closed form defined in (\ref{complex-Hamiltonian}):  
\begin{enumerate}
\item[i)]  there is a domain denoted by $D(K_N)\subset Q(\kappa_n)$,  where an  m-sectorial -- and thus closed -- operator $K_N \equiv K_N(\tau)$ is defined and the associated form coincides with $\kappa_N(\tau)$ (the operator $K_N$ is uniquely determined by the properties in i));
\item[ii)] the  form domain $Q(\kappa_N)$ coincides with the form domain, $Q(H_{I_{N-1,1}}^0)\equiv D((H_{I_{N-1,1}}^0)^{\frac{1}{2}})$,  of $H_{I_{N-1,1}}^0$. 
\end{enumerate}
Hence, we deduce that $D(K_N))\subseteq D((H_{I_{N-1,1}}^0)^{\frac{1}{2}})$. Furthermore, $D(K_N)$ is dense, since $m$-sectorial operators are densely defined. From \cite[Corollary 2.4, p. 323]{K} we get that~$D(H^0_{I_{N-1,1}})\subseteq D(K_N)$; in fact,  this follows from the following statements: 
\begin{enumerate}
\item[1)] the l-h-s of (\ref{complex-Hamiltonian}) is a well defined operator on $D(H^0_{I_{N-1,1}})$ due to the assumption $D((H_{I_{k,i}}^0)^{\frac{1}{2}})\subseteq D(V_{I_{k,i}})$ for any $I_{k,i}$, and the form induced by this operator coincides with $\kappa_{N}(\tau)$ restricted to $D(H^0_{I_{N-1,1}})$; 
\item[2)]
 by   \cite[Corollary 2.4, p. 323]{K}, the operator $K_N$  extends the operator considered in 1).
 \end{enumerate}

The constraint in (\ref{klmn-cond}) readily implies
%\footnote{{\color{blue}Notice that the operator $B_{I_{k,i}}:=(H_{I_{k,i}}^0+1)^{-\frac{1}{2}}V_{I_{k,i}}(H_{I_{k,i}}^0+1)^{-\frac{1}{2}}$ is defined everywhere since $H_{I_{k,i}}^0\geq 0$,  hence any $\psi\in \mathcal{H}^{(N)}$ can be written as $\psi=(H_{I_{k,i}}^0+1)^{\frac{1}{2}}\phi$ for some $\phi \in D((H_{I_{k,i}}^0)^{\frac{1}{2}}) \subset D(V_{I_{k,i}})$. Since $V_{I_{k,i}}$ is symmetric by assumption then $B_{I_{k,i}}$ is symmetric too; consequently,  being bounded and symmetric $B_{I_{k,i}}$ is self-adjoint.  Finally we can estimate $\|B_{I_{k,i}}\|=\sup_{\psi \,;\,\|\psi\|=1}\Big|\langle \psi\,,\,B_{I_{k,i}}\psi \rangle\Big|\leq a$ by writing $\psi =\frac{(H_{I_{k,i}}^0+1)^{\frac{1}{2}}\phi}{\|(H_{I_{k,i}}^0+1)^{\frac{1}{2}}\phi \|}\,$ and using the from bound (\ref{klmn-cond}).}} 
that 
\begin{equation}\label{weighted}
\|(H_{I_{k,i}}^0+1)^{-\frac{1}{2}}V_{I_{k,i}}(H_{I_{k,i}}^0+1)^{-\frac{1}{2}}\|\leq a\,.
\end{equation}
This motivates introducing the weighted norm
\begin{equation}
\|V_{I_{k,i}}\|_{H^0}:=\|(H_{I_{k,i}}^0+1)^{-\frac{1}{2}}V_{I_{k,i}}(H_{I_{k,i}}^0+1)^{-\frac{1}{2}}\|\,.
\end{equation}
Without loss of generality, we may assume that $a=\frac{1}{2}$.

Our results apply to anharmonic quantum crystal models described by Hamiltonians of the form
\begin{equation}
K^{crystal}_{N}:=\sum_{j=1}^{N}\Big(-\frac{d^2}{dx^2_j}+V(x_j))\Big)+t\sum_{j=1}^{N-1}W(x_j,x_{j+1})=:\sum_{j=1}^{N}H_j+t\sum_{j=1}^{N-1}W(x_j,x_{j+1})
\end{equation}
acting on the Hilbert space $\mathcal{H}^N:=\otimes_{j=1}^{N}L^2(\mathbb{R}\,,\,dx_j)$, with $V(x_j)\geq 0$,  $V(x_j)\to \infty$, for $|x_j|\to \infty$,  $D((H_j+H_{j+1})^{\frac{1}{2}})\subseteq D(W(x_j,x_{j+1}))$, and $W(x_j,x_{j+1})$ form-bounded by $H_j+H_{j+1}$.  The class described above includes the $\phi^4-$model on the one-dimensional lattice, corresponding to $V(x_j)=x_j^2+x_j^4$ and $W(x_i,x_j)=x_jx_{j+1}$.

\subsection{Main result}
The main result in this paper is the following theorem proven in Section \ref{block-bound}; see Theorems \ref{main-res} and \ref{analyticity}.\\

{\bf{Theorem.}}
\emph{Under the assumption that (\ref{gaps}), (\ref{potential}) and (\ref{klmn-cond}) hold, the Hamiltonian $K_{N}$ defined in (\ref{complex-Hamiltonian}) has the following properties: For some $t_0 > 0$ and for any $\tau \in \mathbb{C}$ with $\vert \tau \vert < t_0$, there exists a suitable invertible operator $U_{N}(\tau)$ such that  the operator $U^{-1}_{N}(\tau)K_{N}(\tau)U_{N}(\tau)$ has the following properties:
\begin{enumerate}
\item It has a nondegenerate eigenvalue $E_N(\tau)$ analytic in $\tau$ for $\vert \tau \vert < t_0$;
\item the rest of its spectrum is at a distance larger than or equal to $\frac{1}{2}$  from $E_N(\tau)$;
\item for $\tau=t\in \mathbb{R}$, the operator $U_{N}(\tau)$ is unitary,  and  the eigenvalue $E_N(\tau\equiv t)$ is the ground-state energy of $K_{N}(t)$.
\end{enumerate}
}

\noindent
Our proof is based on the block-diagonalization procedure implemented in \cite{DFPR} for these models but with a real coupling constant. We define
\begin{equation}\label{vacuum-proj}
P_{vac}:=\bigotimes_{i=1}^{N} P_{\Omega_i}\,.
\end{equation}
(Note that $P_{vac}$ is the orthogonal projector onto the ground state of the operator $K_{N}(\tau=0)=\sum_{i=1}^{N} H_{i}$.) We construct an invertible operator $U_{N}(\tau)$ acting on $\mathcal{H}^{(N)}$  with the property that, after conjugation, the operator
\begin{equation}\label{conjug}
U^{-1}_{N}(\tau)K_{N}(\tau)U_{N}(\tau)=: \widetilde{K}_{N}(\tau)
\end{equation}
is \textit{``block-diagonal''} with respect to $P_{vac}$, $P_{vac}^{\perp} (:= \charf - P_{vac})$
\begin{equation}\label{block-diag-eq}
\widetilde{K}_{N}(\tau)= P_{vac} \widetilde{K}_{N}(\tau) P_{vac} + P_{vac}^{\perp} \widetilde{K}_{N}(\tau) P_{vac}^{\perp}\,.
\end{equation}
With $E_N(\tau)$ the eigenvalue solving the equation $\widetilde{K}_{N}(\tau)P_{vac}=E_N(\tau)P_{vac}$, we prove that,  for $\vert \tau \vert \leq t_0$,
\begin{equation}\label{gapss}
\text{spec}\left(P_{vac}^{\perp}(\widetilde{K}_{N}(\tau)-E_N(\tau)) P_{vac}^{\perp}  \upharpoonright_{P_{vac}^{\perp} \mathcal{H}^{(N)}}\right)\cap \overline{\mathbb{D}_{\frac{1}{2}}}=\emptyset
\,,
\end{equation}
with $\mathbb{D}_{\rho}:=\{z\in \mathbb{C}\,;\,|z|< \rho\}$, \textit{uniformly} in $N$.

\noindent
From the theorem above we also derive the following result (see Proposition \ref{thermo}):

{\bf{Proposition.}}
\emph{If the potentials $V_{I_{k,i}}$, $k\leq \bar{k}$,  are invariant under translations, then the limiting function $$\varepsilon (\tau):=\lim_{N\to \infty}\frac{E_N(\tau)}{N}$$ exists for any $|\tau|\leq t_0$ and is analytic in $\tau$ for $\vert \tau \vert < t_0$.}
\\

The iterative construction of the operator $U_{N}(\tau)$ coincides with an analogous one for a 
 real coupling constant. For completeness, the scheme of the diagonalization is illustrated in detail  in Sect. \ref{conjugations}.    In Sect. \ref{block-diag}, the proof of convergence of our construction of the operator $U_{N}(\tau)$ is discussed by explaining some of the modifications needed in the complex case,  in particular the proof of the property displayed in (\ref{gapss}) is given.  Some of the proofs are deferred to Appendix A.  The main result of the paper concerning the analyticity of $E_N(\tau)$ is presented in Sect. \ref{block-bound}, Theorem \ref{analyticity}.\\

%\begin{rem}
%We know that ideas developed in this paper can be extended to quantum chains of bosonic degrees of freedom.
%\end{rem}

{\bf{Notation}}
\\

\noindent
1) Notice that $I_{k,q}$  can also be seen as a connected one-dimensional graph with $k$ edges connecting  the $k+1$ vertices $q,1+q,\dots, k+q$, or as an ``interval'' of length $k$ whose left end-point coincides with $q$. 
\\

\noindent
2) We use the same symbol for the operator $O_j$ acting on $\mathcal{H}_j$ and the corresponding operator $$\charf_{i}\otimes\dots \otimes  \charf_{j-1}\otimes O_j \otimes \charf_{j+1}\dots \otimes \charf_l$$ acting on $ \bigotimes_{k=i}^{l} \mathcal{H}_k$, for any $i\leq j\leq l$.
\\

\noindent
3) With the symbol ``$\subset$" we denote strict inclusion, otherwise we use the symbol  ``$\subseteq$". 
\\

{\bf{Acknowledgements.}}
A.P.  thanks  the Pauli Center, Z\"urich, for hospitality in Spring 2017 when this project got started. S. D. V. and S. R. are supported  by the ERC Advanced Grant 669240 QUEST "Quantum Algebraic Structures and Models". S. D.V., A.P., and S. R. also acknowledge the MIUR Excellence Department Project awarded to the Department of Mathematics, University of Rome Tor Vergata, CUP E83C18000100006.

\setcounter{equation}{0}
\section{\emph{Local} conjugations based on Lie-Schwinger series}\label{conjugations}

In this section we describe some of the key ideas underlying our proof of the theorem announced in the previous section.
We study quantum chains with Hamiltonians $K_{N}(\tau)$ associated with the quadratic form described in (\ref{complex-Hamiltonian}) acting on the Hilbert space $\mathcal{H}^{(N)}$ defined in (\ref{tensorprod}).
As explained in Sect. 1, our aim is to block-diagonalize $K_{N}(\tau)$, for $\vert \tau \vert$ small enough, by conjugating it with
 a sequence of operators chosen according to the ``Lie-Schwinger procedure'' (supported on subsets of $\lbrace 1, \dots, N \rbrace$ of successive sites), which for $\tau=t\in \mathbb{R}$ are unitary operators. The block-diagonalization will concern operators acting on tensor-product  spaces of the type
$\mathcal{H}_{q} \otimes \dots \otimes \mathcal{H}_{k+q}$ (and acting trivially on the remaining tensor factors), and it will be with respect to the projection onto the ground-state (``vacuum'') subspace, $\lbrace \mathbb{C}(\Omega_{q}\otimes \dots \otimes \Omega_{k+q})\rbrace$, contained in $\mathcal{H}_{q} \otimes \dots \otimes \mathcal{H}_{k+q}$ and its orthogonal complement. Along the way, new interaction terms are created  whose supports correspond to increasingly longer intervals (connected subsets) of the chain. 

The block-diagonalization procedure for unbounded interactions treated in this paper is formally identical to the scheme introduced in \cite{DFPR}. Hence the formal aspects described in the next section are unchanged w.r.t. \cite{DFPR}.  Nevertheless, the lack of  self-adjointness of the operators $K_N(\tau)$ requires
modifications in the proofs,  in particular in the control of the spectrum of the operators $G_{I_{k,q}}$; see (\ref{def-G}).

\subsection{Block-diagonalization: Definitions and formal aspects}
For each $k$, we consider $(N-k)$ block-diagonalization steps, each of them associated with  a subset $I_{k,q},\, q=1, \dots, N-k$. 
The block-diagonalization of the Hamiltonian will be  with respect to the subspaces associated with the projectors in (\ref{pro-minus})-(\ref{pro-plus}), introduced below.
By $(k,q)$ we label the block-diagonalization step associated with $I_{k,q}$. We introduce an ordering amongst these steps:
\begin{equation}
(k',q') \succ (k,q)
\end{equation} 
if $k'> k$ or if $k'=k$ and $q'>q$. 

\noindent
Our original Hamiltonian is denoted by $K^{(0,N)}_N :=K_{N}(\tau)$. We proceed to the first block-diagonalisation step yielding $K_{N}^{(1,1)}$. The index $(0,N)$  is our initial choice of the index $(k,q)$: all the on-site terms in the Hamiltonian, i.e, the terms $H_i$, are block-diagonal with respect to the subspaces associated with the projectors in (\ref{pro-minus})-(\ref{pro-plus}), for $l=0$.
Our goal is to arrive at a Hamiltonian of the form
\begin{eqnarray}\label{kappa-k-q}
K_N^{(k,q)}
& :=&\sum_{i=1}^{N}H_{i}+\tau \sum_{i=1}^{N-1}V^{(k,q)}_{I_{1,i}}+\tau\sum_{i=1}^{N-2}V^{(k,q)}_{I_{2,i}}+\dots+\tau\sum_{i=1}^{N-k}V^{(k,q)}_{I_{k,i}} \label{def-transf-ham}\\
& &+\tau\sum_{i=1}^{N-k-1}V^{(k,q)}_{I_{k+1,i}}+\dots+\tau\sum_{i=1}^{2}V^{(k,q)}_{I_{N-2,i}}+\tau V^{(k,q)}_{I_{N-1,1}}\label{def-transf-ham-bis}
\end{eqnarray}
(after the block-diagonalization step $(k,q)$)
with the following properties:
\begin{enumerate}
\item
For a fixed $I_{l,i}$,  the corresponding potential term changes, at each step of the block-diagonalization procedure, up to the step $(k,q)\equiv (l,i)$; hence $V^{(k,q)}_{I_{l,i}}$ is the potential term associated with the interval $I_{l,i}$ at step $(k,q)$ of the block-diagonalization, and the superscript $(k,q)$ keeps track of the changes in the potential term in step $(k,q)$. The operator $V^{(k,q)}_{I_{l,i}}$ acts as the identity on the spaces $\mathcal{H}_j$ for $j\neq i,i+1,\dots,i+l$; the description of how these terms are created and estimates on their norms are deferred to Sects \ref{algo} and \ref{block-bound};
\item
for all sets $I_{l,i}$ with $ (l,i)\prec (k,q)$ and for the set $I_{l,i} \equiv I_{k,q} $, the associated potential $V^{(k,q)}_{I_{l, i}}$ is block-diagonal w.r.t. the decomposition of the identity into the sum of projectors
\begin{equation}\label{pro-minus}
P^{(-)}_{I_{l,i}}:= P_{\Omega_{i}}\otimes P_{\Omega_{i+1}}\otimes \dots \otimes P_{\Omega_{i+l}}\,,
\end{equation}
\begin{equation}\label{pro-plus}
P^{(+)}_{I_{l,i}}:= (P_{\Omega_{i}}\otimes P_{\Omega_{i+1}}\otimes \dots \otimes P_{\Omega_{i+l}})^{\perp}\,.
\end{equation}
\end{enumerate}
%\item
%from some $k-$dependent $l$ on, all the interaction terms $V^{(k,q)}_{I_{l,i}}$ may vanish.
\begin{rem}
We warn the reader that new potentials created along the block-diagonalization process are $\tau$-dependent though this is not reflected in our notation.
\end{rem}

\begin{rem}\label{remark-decomp}
It is important to notice that if $V^{(k,q)}_{I_{l,i}}$ is block-diagonal w.r.t. the decomposition of the identity into $$P^{(+)}_{I_{l,i}}+P^{(-)}_{I_{l,i}}\,,$$
i.e., $$V^{(k,q)}_{I_{l,i}}=P^{(+)}_{I_{l,i}}V^{(k,q)}_{I_{l,i}}P^{(+)}_{I_{l,i}}+P^{(-)}_{I_{l,i}}V^{(k,q)}_{I_{l,i}}P^{(-)}_{I_{l,i}}\,\,,$$  then, for $ I_{l,i} \subset I_{r,j}$,
we have that
$$P^{(+)}_{I_{r,j}}\Big[P^{(+)}_{I_{l,i}}V^{(k,q)}_{I_{l,i}}P^{(+)}_{I_{l,i}}+P^{(-)}_{I_{l,i}}V^{(k,q)}_{I_{l,i}}P^{(-)}_{I_{l,i}}\Big]P^{(-)}_{I_{r,j}}=0\,.$$
To see that the first term vanishes, we use that
\begin{equation}
P^{(+)}_{I_{l,i}}\,P^{(-)}_{I_{r,j}}=0\,,
\end{equation}
while, in the second term, we use that
\begin{equation}
P^{(-)}_{I_{l,i}}V^{(k,q)}_{I_{l,i}}P^{(-)}_{I_{l,i}}\,P^{(-)}_{I_{r,j}}=P^{(-)}_{I_{r,j}}P^{(-)}_{I_{l,i}}V^{(k,q)}_{I_{l,i}}P^{(-)}_{I_{l,i}}P^{(-)}_{I_{r,j}}
\end{equation}
and
\begin{equation}
P^{(+)}_{I_{r,j}}P^{(-)}_{I_{r,j}}=0\,.
\end{equation}

\noindent
Hence $V^{(k,q)}_{I_{l,i}}$ is also block-diagonal with respect to the decomposition of the identity into  $$P^{(+)}_{I_{r,j}}+P^{(-)}_{I_{r,j}}\,.$$
Note, however, that
\begin{equation}
P^{(-)}_{I_{r,j}}\Big[P^{(+)}_{I_{l,i}}V^{(k,q)}_{I_{l,i}}P^{(+)}_{I_{l,i}}+P^{(-)}_{I_{l,i}}V^{(k,q)}_{I_{l,i}}P^{(-)}_{I_{l,i}}\Big]P^{(-)}_{I_{r,j}}
=P^{(-)}_{I_{r,j}}\,V^{(k,q)}_{I_{l,i}}\,P^{(-)}_{I_{r,j}} \,.
\end{equation}
But 
$$P^{(+)}_{I_{r,j}}\Big[P^{(+)}_{I_{l,i}}V^{(k,q)}_{I_{l,i}}P^{(+)}_{I_{l,i}}+P^{(-)}_{I_{l,i}}V^{(k,q)}_{I_{l,i}}P^{(-)}_{I_{l,i}}\Big]P^{(+)}_{I_{r,j}}$$
remains as it is.
\end{rem}
\begin{rem}\label{rem-block}
The block-diagonalization procedure that we will implement enjoys the property that the terms  block-diagonalized along the process  do not change, anymore, in subsequent steps.\\

\end{rem}
\subsection{Lie-Schwinger conjugation associated with $I_{k,q}$}\label{L-S-scheme}
Here we explain the block-diagonalization procedure from $(k,q-1)$ to $(k,q)$ by which the term $V^{(k,q-1)}_{I_{k,q}}$ is transformed  to a new operator, $V^{(k,q)}_{I_{k,q}}$, that is block-diagonal  w.r.t.  the decomposition of the identity into  $$P^{(+)}_{I_{k,q}}+P^{(-)}_{I_{k,q}}\,.$$
We note that,   because the first index (i.e., the number of edges of the interval) is changing from $k$ to $k+1$, the steps $(k, N-k)\, \rightarrow \,(k+1, 1)$ are somewhat different\footnote{The initial step, $(0,N)\rightarrow (1,1)$,  is of this type; see the definitions in (\ref{initial-V}) of the terms in the Hamiltonian $K_N$ with \emph{nearest-neighbor interactions}.}. Here we deal with general steps $(k, q-1)\, \rightarrow \,(k, q)$, with $N-k\geq q\geq 2$, and we refer the reader to \cite{DFPR} for the special steps mentioned above that require a slightly different notation.

\noindent
We recall that the Hamiltonian $K_N^{(k,q-1)}$ is given by
\begin{eqnarray}
K_N^{(k,q-1)}
& :=&\sum_{i=1}^{N}H_{i}+\tau\sum_{i=1}^{N-1}V^{(k,q-1)}_{I_{1,i}}+\tau\sum_{i=1}^{N-2}V^{(k,q-1)}_{I_{2,i}}+\dots+\tau\sum_{i=1}^{N-k}V^{(k,q-1)}_{I_{k,i}} \\
& &+\tau\sum_{i=1}^{N-k-1}V^{(k,q-1)}_{I_{k+1,i}}+\dots+\tau\sum_{i=1}^{2}V^{(k,q-1)}_{I_{N-2,i}}+\tau V^{(k,q-1)}_{I_{N-1,1}}
\end{eqnarray}
and has the following properties
\begin{enumerate}
\item
each operator $V^{(k,q-1)}_{I_{l,i}}$ acts as the identity on the spaces $\mathcal{H}_j$ for $j\neq i,i+1,\dots,i+l$. In Sect. \ref{algo} we explain how these terms are created, and in Sect. \ref{block-bound} how their norms can be estimated;
\item
each operator $V^{(k,q-1)}_{I_{l,i}}$, with $l<k$ or $l=k$ and $q-1\geq i$,  is block-diagonal w.r.t. the decomposition of the identity into the sum of projectors in (\ref{pro-minus})-(\ref{pro-plus}).
%\begin{equation}\label{pro-minus-bis}
%P^{(-)}_{I_{l,i}}:=P_{\Omega_{i}}\otimes P_{\Omega_{i+1}}\otimes \dots \otimes P_{\Omega_{i+l}}\,,
%\end{equation}
%\begin{equation}\label{pro-plus-bis}
%P^{(+)}_{I_{l,i}}:= (P_{\Omega_{i}}\otimes P_{\Omega_{i+1}}\otimes \dots \otimes P_{\Omega_{i+l}})^{\perp}\,.
%\end{equation}
%\item
%from some $k-$dependent $l$ on, the interaction $V^{(k,N-k)}_{I_{l,i}}$ may be all zero.
\end{enumerate}
\begin{rem}
The term \emph{step} is used throughout the paper with two slightly different meanings: 
\begin{enumerate}
\item[i)] as \emph{level} in the block-diagonalization iteration, e.g., $K_N^{(k,q)}$ is the Hamiltonian in step $(k,q)$; 
\item[ii)] for the block-diagonalization procedure to switch from level $(k,q-1)$ to level $(k,q)$, e.g., the step $(k,q-1)\rightarrow (k,q)$.
\end{enumerate}
\end{rem}

With the next block-diagonalization step, labeled by $(k,q)$, we want to block-diagonalize the interaction term $V^{(k,q-1)}_{I_{k,q}}$, considering the operator
\begin{equation}\label{def-G}
G_{I_{k,q}}:=\sum_{i\subset I_{k,q} }H_i+\tau \sum_{I_{1,i} \subset I_{k,q}} V^{(k, q-1)}_{I_{1,i}}+\dots+\tau \sum_{I_{k-1,i}\subset I_{k,q}}V^{(k, q-1)}_{I_{k-1,i}}\,,
\end{equation}
as the ``unperturbed" Hamiltonian. This operator is block-diagonal w.r.t. the decomposition of  the identity, i.e., 
\begin{equation}
G_{I_{k,q}}=P^{(+)}_{I_{k,q}}G_{I_{k,q}}P^{(+)}_{I_{k,q}}+P^{(-)}_{I_{k,q}}G_{I_{k,q}}P^{(-)}_{I_{k,q}}\,;
\end{equation}
see Remarks \ref{remark-decomp} and \ref{rem-block}.
We also define
\begin{equation} \label{def-E-bis}
E_{I_{k,q}}:=\langle \Omega_{q}\otimes \Omega_{2}\otimes \dots \otimes \Omega_{k+q}\,,\, G_{I_{k,q}} \Omega_{q}\otimes \Omega_{2}\otimes \dots \otimes \Omega_{k+q} \rangle \,
\end{equation}
so that $$G_{I_{k,q}}P^{(-)}_{I_{k,q}}=E_{I_{k,q}}P^{(-)}_{I_{k,q}}\,.$$
 Next, we sketch a convenient formalism used to construct our block-diagonalisation operations; for further details the reader is referred to Sects. 2 and 3 of \cite{DFFR}. For operators $A$ and $B$, we define
\begin{equation}
ad\, A\,(B):=[A\,,\,B]\,,
\end{equation}
and, for $n\geq 2$,
\begin{equation}
ad^n A\,(B):=[A\,,\,ad^{n-1} A\,(B)]\,.
\end{equation}
 This definition is in general only formal for unbounded operators; in fact, the $B$ operators in the formulae below are unbounded.  But, as shown in the proof of Theorem \ref{norms}, the formula is still meaningful for the operators considered in this paper.
In the block-diagonalization step $(k,q)$, we use the operator
\begin{equation}
S_{I_{k,q}}:=\sum_{j=1}^{\infty}\tau^j(S_{I_{k,q}})_j\,,
\end{equation}
where the terms $(S_{I_{k,q}})_j$ are defined iteratively; (notice that our definition is meaningful, since $(V^{(k,q-1)}_{I_{k,q}})_j$ depends on the operators $(V^{(k,q-1)}_{I_{k,q}})_1$ and $(S_{I_{k,q}})_r$,  with $r<j$):
\begin{itemize}
\item
\begin{eqnarray}\label{def-S-bis}
(S_{I_{k,q}})_j
%&:=&ad^{-1}\,G_{I_{k,q}}\,((V^{(k, N-k)}_{I_{k,q}})^{od}_j)\\
&:=&\frac{1}{G_{I_{k,q}}-E_{I_{k,q}}}P^{(+)}_{I_{k,q}}\,(V^{(k, N-k)}_{I_{k,q}})_j\,P^{(-)}_{I_{k,q}}\\
& &-P^{(-)}_{I_{k,q}}(V^{(k, N-k)}_{I_{k,q}})_j\,P^{(+)}_{I_{k,q}}\frac{1}{G_{I_{k,q}}-E_{I_{k,q}}}\,,
\end{eqnarray}
\item
$$(V^{(k,q-1)}_{I_{k,q}})_1=V^{(k,q-1)}_{I_{k,q}}\,,$$ 
and, for $j\geq 2$,\\
\vspace{0.2cm}
$(V^{(k,q-1)}_{I_{k,q}})_j\,:=$
\begin{eqnarray}
\quad& =&\sum_{p\geq 2, r_1\geq 1 \dots, r_p\geq 1\, ; \, r_1+\dots+r_p=j}\frac{1}{p!}\text{ad}\,(S_{I_{k,q}})_{r_1}\Big(\text{ad}\,(S_{I_{k,q}})_{r_2}\dots (\text{ad}\,(S_{I_{k,q}})_{r_p}(G_{I_{k,q}})\dots \Big)\nonumber \\
&+&\sum_{p\geq 1, r_1\geq 1 \dots, r_p\geq 1\, ; \, r_1+\dots+r_p=j-1}\frac{1}{p!}\text{ad}\,(S_{I_{k,q}})_{r_1}\Big(\text{ad}\,(S_{I_{k,q}})_{r_2}\dots (\text{ad}\,(S_{I_{k,q}})_{r_p}(V^{(k,q-1)}_{I_{k,q}})\dots \Big)\, . \nonumber\\
\end{eqnarray}
\end{itemize}
The operator $S_{I_{k,q}}$ will turn out to be bounded and, consequently,  $e^{S_{I_{k,q}}}$ is invertible. We will prove that
\begin{equation}\label{def-KN}
K_N^{(k,q)}=e^{S_{I_{k,q}}}\,K_N^{(k, q-1)}\,e^{-S_{I_{k,q}}}\,
\end{equation}
where the l-h-s in (\ref{def-KN}) involves the effective potentials $V^{(k,q)}_{l,i}$ (see Sect. \ref{algo}) defined in such a way that,  \emph{a posteriori}, the identity above holds.
\\

\begin{rem}\label{V&S}
The formal sums defining the operators $(V^{(k,q-1)}_{I_{k,q}})_j$ and $(S_{I_{k,q}})_j$, and the series defining $V^{(k,q)}_{I_{k,q}}$ and $S_{I_{k,q}}$,  are  controlled similarly to the real coupling constant case treated in \cite{DFPR}. This is the content of Sect. \ref{block-diag}, with some of the proofs deferred to the Appendix,
where the operators $(V^{(k,q-1)}_{I_{k,q}})_j$ and $V^{(k,q)}_{I_{k,q}}$ will be shown to be bounded in the norm $\|\cdot \|_{H^0}$, whereas the operator $S_{I_{k,q}}$  will turn out to be bounded.  The more regular behaviour of $S_{I_{k,q}}$ is due to the projectors entering the definition of $(S_{I_{k,q}})_j$, since one of them, $P^{(-)}_{I_{k,q}}$, is of finite rank.
\end{rem}
\subsection{The algorithm $\alpha_{I_{k,q}}$}\label{algo}
The interaction terms arising in  our block-diagonalization steps are controlled by an algorithm, $\alpha_{I_{k,q}}$, which determins a map that sends each operator $V^{(k,q-1)}_{I_{l,i}}$ to a corresponding potential term supported on the same interval, but at the next block-diagonalization step, i.e., 

\begin{equation}
\alpha_{I_{k,q}}(V^{(k,q-1)}_{I_{l,i}})=:V^{(k,q)}_{I_{l,i}}\,.
\end{equation}

We start from $V_{I_{0,i}}^{(0,N)}:=H_i$ and follow
the evolution of these operators as well as that  of the potential terms.  In Definition \ref{def-interections}, we present  the \textit{iterative definition} of the operators $$V^{(k,q)}_{I_{l,i}}:=\alpha_{I_{k,q}}(V^{(k,q-1)}_{I_{l,i}})$$ in terms of the operators, $V^{(k,q-1)}_{I_{l,i}}$, at the previous step $(k,q-1)$, starting from 
\begin{equation}\label{initial-V}
V_{I_{0,i}}^{(0,N)}\equiv H_i\quad ,\quad
V_{I_{1,i}}^{(0,N)}\equiv V_{I_{1,i}}\quad,\quad
V_{I_{l,i}}^{(0,N)} =0\,\,\text{for}\,\, l\geq 2\,.
\end{equation}

We warn the reader that the definitions below involve unbounded operators; see Remark \ref{unbound}, below.
\begin{defn}\label{def-interections}
We assume that, for fixed $(k,q-1)$, with $(k,q-1) \succ (0,N)$, the operators $V^{(k,q-1)}_{I_{l,i}}$ and $S_{I_{k,q}}$ are well defined, for any $l,i$; or  we assume that $(k,q)=(1,1)$ and that the operator $S_{I_{1,1}}$ is well defined. We then define the operators $V^{(k,q)}_{I_{l,i}}$ as follows (but note that  if $q=1$ the couple $(k,q-1)$ is replaced by $(k-1,N-k+1)$ in (\ref{case-in})-(\ref{main-def-V-bis})); see Fig. \ref{fig:cases-bis} for a graphical representation of the different cases b), c) d-1) and d-2, below:
\begin{figure}
 \includegraphics[width=\linewidth]{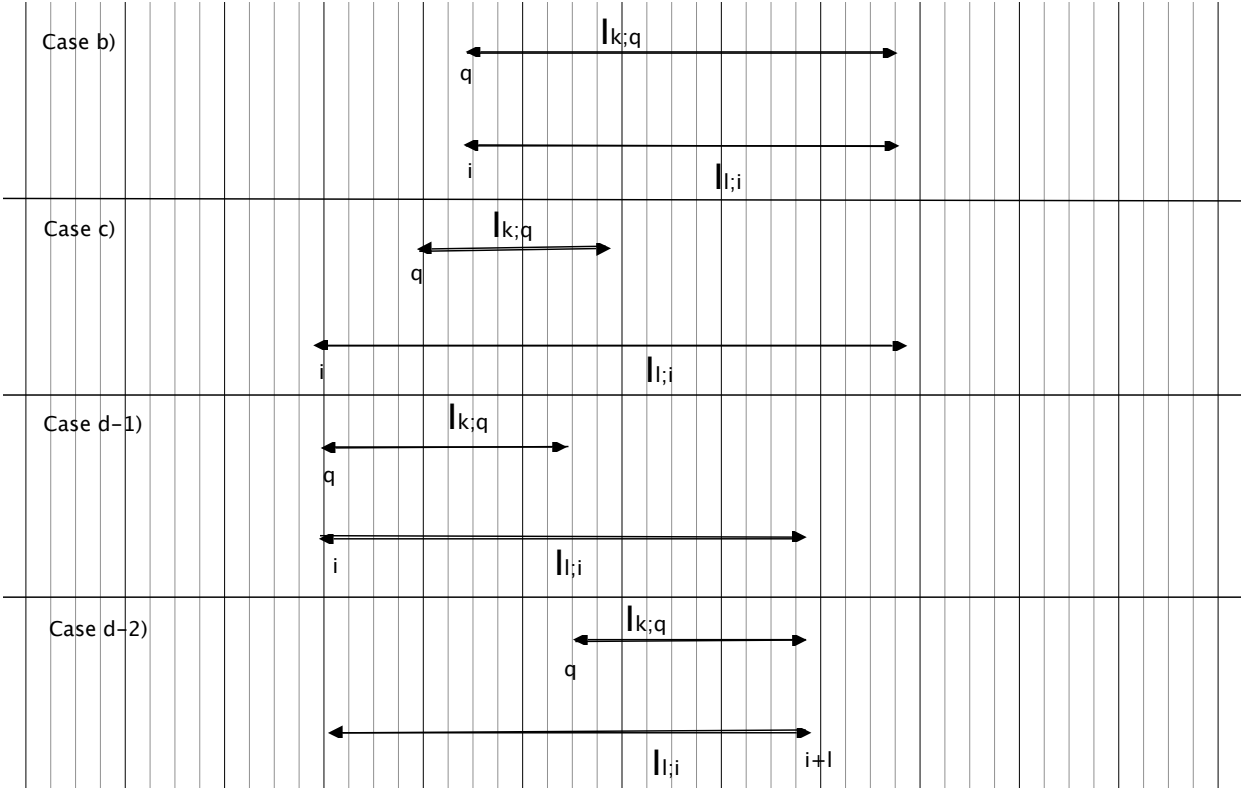}
 \caption{Relative positions of intervals $I_{k,q}$ and $I_{l,i}$}
 \label{fig:cases-bis}
\end{figure}
\begin{itemize}
\item[a)]
in all the following cases
\begin{itemize}
\item[a-i)] $l\leq k-1$;
\item[a-ii)]  $I_{l,i}\cap I_{k,q}=\emptyset$;
\item[a-iii)]  $ I_{l,i}\cap I_{k,q}\neq\emptyset$ but $l\geq k$ and $I_{k,q} \nsubseteq I_{l,i}$;
\end{itemize}
we define
\begin{equation}\label{case-in}
V^{(k,q)}_{I_{l,i}}:=V^{(k,q-1)}_{I_{l,i}}\,;
\end{equation}

\item[b)]
if $I_{l,i}\equiv I_{k,q}$, we define
\begin{equation}
V^{(k,q)}_{I_{l,i}}:= \sum_{j=1}^{\infty}\tau^{j-1}(V^{(k,q-1)}_{I_{l,i}})^{diag}_j \,;
\end{equation}
\item[c)]
 if $I_{k,q}\subset I_{l,i}$ and $i, i+l \notin I_{k,q}$, we define
\begin{equation}
V^{(k,q)}_{I_{l,i}}:=V^{(k,q-1)}_{I_{l,i}}\,+\sum_{n=1}^{\infty}\frac{1}{n!}\,ad^{n}S_{I_{k,i}}(V^{(k,q-1)}_{I_{l,i}})\,;
\end{equation}

\item[d)]

\noindent
if $I_{k,q}\subset I_{l,i}$ and either $i$ or $i+l$ belongs to $ I_{k,q}$, we define
\begin{itemize}
\item[ d-1)] if $i$ belongs to $ I_{k,q}$, i.e., $q \equiv i$,  then
\begin{eqnarray}
V^{(k,q)}_{I_{l,i}} &:= &V^{(k,q-1)}_{I_{l,i}}\,+\sum_{j=0}^{k}\sum_{n=1}^{\infty}\frac{1}{n!}\,ad^{n}S_{I_{k,i}}(V^{(k,q-1)}_{I_{l-j,i+j}})\,; \label{main-def-V}
\end{eqnarray}

\item[ d-2)] if $i+l$ belongs to $ I_{k,q}$, i.e., $q +k\equiv i+l$ that means $q\equiv i+l-k$,  then
\begin{eqnarray}
V^{(k,q)}_{I_{l,i}} &:= & V^{(k,q-1)}_{I_{l,i}}\,+\sum_{j=0}^{k}\sum_{n=1}^{\infty}\frac{1}{n!}\,ad^{n}S_{I_{k;i+l-k}}(V^{(k,q-1)}_{I_{l-j,i}})\,. \label{main-def-V-bis}
\end{eqnarray}
\end{itemize}
Notice that  in both cases, d-1) and d-2), the elements of the sets $\{I_{l-j,i+j}\}_{j=1}^{k}$ and $\{I_{l-j,i}\}_{j=1}^{k}$, respectively,  are all the intervals, $\mathscr{I}$, such that $\mathscr{I} \cap I_{k,q} \neq \emptyset $,  $\mathscr{I} \nsubseteq I_{k,q}$,  $I_{k,q}\nsubseteq \mathscr{I} $, and   $\mathscr{I} \cup I_{k,q}\equiv I_{l,i}$.
\end{itemize}
\end{defn}

\begin{rem}\label{unbound}
The results in Theorems \ref{norms} and  the argument in (\ref{domain-in})-(\ref{domain-fin}), below,  imply that
the quantities defined in (\ref{case-in})-(\ref{main-def-V-bis}) are to be understood as  quadratic forms on the domain $D((H^0_{I_{l,i}})^{\frac{1}{2}})$.
\end{rem}

\begin{rem}
Notice that, according to  Definition \ref{def-interections}:
\begin{itemize}
\item if $(k', q')\succ (l,i)$  then
\begin{equation}
V^{(k',q')}_{I_{l,i}}=V^{(l,i)}_{I_{l,i}}\,,
\end{equation}
since cases b), c), d-1), and d-2) do not arise;
\item
 for $k\geq 1$ and all  allowed choices of $q$, 
\begin{equation}
V^{(k,q)}_{I_{0,i}}=H_i\,,
\end{equation}
due to a-i);
\item
the following identity holds
\begin{equation}\label{conjugation}
e^{S_{I_{k,q}}}\,(G_{I_{k,q}}+\tau V^{(k,q-1)}_{I_{k,q}})\,e^{-S_{I_{k,q}}}=G_{I_{k,q}}+\tau \sum_{j=1}^{\infty}\tau ^{j-1}(V^{(k,q-1)}_{I_{k,q}})^{diag}_j\,.
\end{equation}
Hence the \underline{net result} of the conjugation  of the sum of the operators $V^{(k,q-1)}_{I_{l,i}}$  appearing on the left side of  (\ref{conjugation})
can be re-interpreted as follows:

\noindent
a) The operators $V^{(k,q-1)}_{I_{l,i}}$, with $I_{l,i}\subset I_{k,q}$\,, are kept fixed in the step $(k,q-1)\rightarrow (k,q)$, i.e., we define $V^{(k,q)}_{I_{l,i}}:=V^{(k,q-1)}_{I_{l,i}}$,  hence
\begin{eqnarray}G_{I_{k,q}}&=&\sum_{i\subset I_{k,q} }H_i+\tau \sum_{I_{1,i} \subset I_{k,q}} V^{(k, q-1)}_{I_{1,i}}+\dots+\tau \sum_{I_{k-1,i}\subset I_{k,q}}V^{(k, q-1)}_{I_{k-1,i}}\\
&= &\sum_{i\subset I_{k,q} }H_i+\tau \sum_{I_{1,i} \subset I_{k,q}} V^{(k, q)}_{I_{1,i}}+\dots+\tau \sum_{I_{k-1,i}\subset I_{k,q}}V^{(k, q)}_{I_{k-1,i}}\,;
\end{eqnarray}

\noindent
b)  the operator $V^{(k,q-1)}_{I_{k,q}}$ is transformed to the operator $$V^{(k,q)}_{I_{k,q}}:=\sum_{j=1}^{\infty}\tau ^{j-1}(V^{(k,q-1)}_{I_{k,q}})^{diag}_j\,,$$  which is block-diagonal, and $$\|V^{(k,q)}_{I_{k,q}}\|_{H^0}\leq 2\|V^{(k,q-1)}_{I_{k,q}}\|_{H^0}\,,$$ as will be shown, assuming that $|\tau |>0$ is sufficiently small;
\item
the expansion of $e^{S_{I_{k,i}}}V^{(k,q-1)}_{I_{l,i}}e^{-S_{I_{k,i}}}$ in cases c), d-1), and d-2) is controlled (see Theorem \ref{norms}) by exploiting the boundedness of $S_{I_{k,i}}$ and the bound
on the weighted operator norm $\|V^{(k,q-1)}_{I_{l,i}}\|_{H^0}$, which are  proved iteratively by combining  Lemma \ref{unboundedlemmaA3} with Theorem \ref{norms}.
\end{itemize}
\end{rem}

\newpage

\setcounter{equation}{0}

\section{Block-diagonalization of $K_{N}(\tau)$ and analiticity of $E_{N}(\tau)$}\label{block-diag}
In this section we add mathematical rigour to the block-diagonalization procedure described in Sect. \ref{conjugations}, and we prove the main result of the paper concerning the analyticity of $E_{N}(\tau)$. The section is divided into three parts. In Sect. \ref{block-spec} we study the modifications that are needed  (due to the complex coupling constant) to show that $G_{I_{k,q}}$ does not have spectrum in a certain punctured small disk centred at $E_{I_{k,q}}$; in Sect. \ref{block-bound} we outline the control of the weighted norm of the effective potentials, (the proofs are deferred to the appendix); in Sect. \ref{block-analyticity} we state and prove our main result, namely Theorem \ref{analyticity}.

\subsection{Block-diagonalization: Spectrum of the local Hamiltonian $G_{I_{k,q}}$ around $E_{I_{k,q}}$}\label{block-spec}
To simplify our presentation, we consider a \emph{nearest-neighbor interaction} with 
$$\|V_{I_{1,i}}\|_{H^0}:=\|(H_{I_{1,i}}^0+1)^{-\frac{1}{2}}V_{I_{1,i}}(H_{I_{1,i}}^0+1)^{-\frac{1}{2}}\|=\frac{1}{2}\,.$$ 
However, with obvious modifications, our proof can be adapted to general Hamiltonians of the type as in (\ref{Hamiltonian}). Furthermore, we define
\begin{eqnarray}
& &\langle  V^{(k, q-1)}_{I_{j,i}} \rangle\\
 &:= &\langle \Omega_{i}\otimes \Omega_{i+1}\dots\otimes \Omega_{i+j} \,,\,V^{(k,q-1)}_{I_{j,i}} \,\Omega_{i}\otimes \Omega_{i+1}\dots\otimes \Omega_{i+j} \rangle\\
&=&\langle \Omega_{i}\otimes \Omega_{i+1}\dots\otimes \Omega_{i+j}\,,\,(H_{I_{j,i}}^0+1)^{-\frac{1}{2}}V^{(k,q-1)}_{I_{j,i}} (H_{I_{j,i}}^0+1)^{-\frac{1}{2}}\,\Omega_{i}\otimes \Omega_{i+1}\dots\otimes \Omega_{i+j}\rangle\,,\quad\quad \label{vac-exp}
\end{eqnarray}
which, using the definition in (\ref{def-E-bis}), implies the following identity
\begin{equation}
E_{I_{k,q}}=\tau \Big\{\sum_{I_{1,i} \subset I_{k,q}} \langle V^{(k,q-1)}_{I_{1,i}}\rangle+\dots+\sum_{I_{k-1,i} \subset I_{k,q}} \langle V^{(k,q-1)}_{I_{k-1,i}}\rangle\Big\}\,.
\end{equation}

\noindent
Our induction hypothesis is that, for  $|\tau|>0$ small enough, and for arbitrary $(l,i)$,
\begin{equation}\label{ass-2}
\|(H_{I_{l,i}}^0+1)^{-\frac{1}{2}}V^{(k,q-1)}_{I_{l,i}}(H_{I_{l,i}}^0+1)^{-\frac{1}{2}}\|=:\|V^{(k,q-1)}_{I_{l,i}}\|_{H^0} \leq |\tau|^{\frac{l-1}{4}}\,.
\end{equation}  

\begin{rem}\emph{(Domain of $G_{I_{k,q}}$)}
Assuming the bound in (\ref{ass-2}), the formal expression $G_{I_{k,q}}$ is a well-defined  closed quadratic form, which we denote by $g_{I_{k,q}}$, on the domain $D((H_{I_{k,q}}^0)^{\frac{1}{2}})$. Hence, as in the definition of the operator associated to the closed quadratic form in  (\ref{complex-Hamiltonian}), we can state that, for $|\tau|$ sufficiently small but independent of $k$, $q$, and $N$:  
\begin{enumerate}
\item[i)]  there is a domain that we call  $D(G_{I_{k,q}})\subset Q(g_{I_{k,q}})$ where an  m-sectorial -- and thus closed -- operator $G_{I_{k,q}} \equiv G_{I_{k,q}}(\tau)$ is defined and the associated form coincides with $g_{I_{k,q}}$ (the operator $G_{I_{k,q}}$ is uniquely determined by the properties in i)); 
\item[ii)] the  form domain $Q(G_{I_{k,q}})$ coincides with the form domain, $Q(H_{I_{k,q}}^0)\equiv D((H_{I_{k,q}}^0)^{\frac{1}{2}})$,  of $H_{I_{k,q}}^0$. 
\end{enumerate}
We refer the reader to Theorems 3.9 and 2.1 of \cite{K}. 
\end{rem}

According to the scheme described in Sect. \ref{L-S-scheme},  the operators $V^{(k,q-1)}_{I_{j,i}}$ are block-diagonalized, for arbitrary  $1\leq j \leq k-1$; i.e., 
\begin{equation}\label{informal-in}
V^{(k,q-1)}_{I_{j,i}}=P^{(+)}_{I_{1,i}}V^{(k,q-1)}_{I_{j,i}}P^{(+)}_{I_{1,i}}+P^{(-)}_{I_{1,i}}V^{(k,q-1)}_{I_{j,i}}P^{(-)}_{I_{j,i}}\,.
\end{equation}
Hence  we can  write
\begin{eqnarray}
& &P^{(+)}_{I_{k,q}}\,\Big[\sum_{i\subset I_{k;q;} }H_i+\tau \sum_{I_{1,i} \subset I_{k,q}} V^{(k,q-1)}_{I_{1,i}}+\dots ++\tau \sum_{I_{k-1,i} \subset I_{k,q}} V^{(k,q-1)}_{I_{k-1,i}}\Big]P^{(+)}_{I_{k,q}}\\
&=&P^{(+)}_{I_{k,q}}\,\Big[\sum_{i\subset I_{k,q} }H_i\\
& &\quad\quad+\tau \sum_{I_{1,i} \subset I_{k,q}} P^{(+)}_{I_{1,i}}V^{(k,q-1)}_{I_{1,i}}P^{(+)}_{I_{1,i}}+\tau \sum_{I_{1,i} \subset I_{k,q}} P^{(-)}_{I_{1,i}}V^{(k,q-1)}_{I_{1,i}}P^{(-)}_{I_{1,i}}+\\
& &\quad \quad \dots \\
& &\quad \quad +\tau \sum_{I_{k-1,i} \subset I_{k,q}} P^{(+)}_{I_{k-1,i}}V^{(k,q-1)}_{I_{k-1,i}}P^{(+)}_{I_{1,i}}+\tau \sum_{I_{k-1,i} \subset I_{k,q}} P^{(-)}_{I_{k-1,i}}V^{(k,q-1)}_{I_{k-1,i}}P^{(-)}_{I_{k-1,i}}
\Big]P^{(+)}_{I_{k,q}}\,.\quad\quad\quad \label{1.57}
\end{eqnarray}

\noindent
Recalling that $P^{(-)}_{I_{j,i}}=\charf - P^{(+)}_{I_{j,i}}$, we observe that
\begin{equation}
P^{(-)}_{I_{j,i}}V^{(k,q-1)}_{I_{j,i}}P^{(-)}_{I_{j,i}}=\langle  V^{(k, q-1)}_{I_{j,i}} \rangle P^{(-)}_{I_{j,i}}= \langle V^{(k, q-1)}_{I_{j,i}}\rangle -\langle  V^{(k, q-1)}_{I_{j,i}} \rangle P^{(+)}_{I_{j,i}}\,,
\end{equation}
and, from (\ref{vac-exp}),
\begin{equation}\label{ass-2-bis}
|\langle  V^{(k, q-1)}_{I_{j,i}} \rangle|\leq \|(H_{I_{j,i}}^0+1)^{-\frac{1}{2}}V^{(k,q-1)}_{I_{j,i}} (H_{I_{j,i}}^0+1)^{-\frac{1}{2}}\|\,.
\end{equation}

Next, we define
\begin{eqnarray}
\tau \mathcal{V}_{I_{k,q}}&:=&\tau \sum_{I_{1,i} \subset I_{k,q}} P^{(+)}_{I_{1,i}}(V^{(k,q-1)}_{I_{1,i}}-\langle V^{(k,q-1)}_{I_{1,i}} \rangle) P^{(+)}_{I_{1,i}}+\\
& &+\dots\\
& &+\tau \sum_{I_{k-1,i} \subset I_{k,q}} P^{(+)}_{I_{k-1,i}}(V^{(k,q-1)}_{I_{k-1,i}}-\langle V^{(k,q-1)}_{I_{k-1,i}} \rangle) P^{(+)}_{I_{k-1,i}}\,
\end{eqnarray}
Consequently, we have that
\begin{eqnarray}
& &P^{(+)}_{I_{k,q}}(G_{I_{k,q}}-E_{I_{k,q}})P^{(+)}_{I_{k,q}}\\
&=&P^{(+)}_{I_{k,q}}\,\Big[\sum_{i\subset I_{k,q} }H_i\\
& &\quad\quad+\tau \sum_{I_{1,i} \subset I_{k,q}} P^{(+)}_{I_{1,i}}(V^{(k,q-1)}_{I_{1,i}}-\langle V^{(k,q-1)}_{I_{1,i}} \rangle) P^{(+)}_{I_{1,i}}+\\
& &\quad \quad \dots \\
& &\quad \quad+\tau \sum_{I_{k-1,i} \subset I_{k,q}} P^{(+)}_{I_{k-1,i}}(V^{(k,q-1)}_{I_{k-1,i}}-\langle V^{(k,q-1)}_{I_{k-1,i}}\rangle) P^{(+)}_{I_{k-1,i}}
\Big]P^{(+)}_{I_{k,q}}\,.\quad\quad\quad\\
&=&P^{(+)}_{I_{k,q}}\,\Big[\sum_{i\subset I_{k,q} }H_i\,+\,\tau \mathcal{V}_{I_{k,q}}\Big]P^{(+)}_{I_{k,q}}\,,
\end{eqnarray}
which is a closed operator on $P^{(+)}_{I_{k,q}}D(G_{I_{k,q}})$.
\begin{lem}\label{gap}
Assuming condition (\ref{ass-2}), and choosing $|\tau|$ so small that
\begin{equation}
1-8\tau\,\sum_{j=1}^{+\infty}(j+1)\,|\tau|^{\frac{j-1}{4}}>0\,,
\end{equation}
the following inequality holds  true
\begin{eqnarray}
\|\frac{1}{P^{(+)}_{I_{k,q}}(G_{I_{k,q}}-E_{I_{k,q}}-z)P^{(+)}_{I_{k,q}}}P^{(+)}_{I_{k,q}}\|
&\leq&\frac{2}{1-8|\tau|\,\sum_{j=1}^{+\infty}(j+1)\,|\tau|^{\frac{j-1}{4}}}\,.\label{final-eq-1}
\end{eqnarray}
for arbitrary $z$ with $|z|\leq \frac{1}{2}$.
\end{lem}

\noindent
\emph{Proof.}

\noindent
We propose to use the following expansion, for $|\tau|$ sufficiently small and $z\in \overline{\mathbb{D}_{\frac{1}{2}}}$:
\begin{eqnarray}
& &\frac{1}{P^{(+)}_{I_{k,q}}(G_{I_{k,q}}-E_{I_{k,q}}-z)P^{(+)}_{I_{k,q}}}P^{(+)}_{I_{k,q}}\label{exp-G-in}\\
&=&(\frac{1}{P^{(+)}_{I_{k,q}}\,(\sum_{i\subset I_{k,q} }H_i-z)\,P^{(+)}_{I_{k,q}}})^{\frac{1}{2}}\times \\
& &\times \sum_{l=0}^{\infty}\,\Big\{(\frac{1}{P^{(+)}_{I_{k,q}}\,(\sum_{i \subset I_{k,q} }H_i-z)\,P^{(+)}_{I_{k,q}}})^{\frac{1}{2}}\,[P^{(+)}_{I_{k, q}}\,\tau \mathcal{V}_{I_{k,q}}\,P^{(+)}_{I_{k,q}}]\,(\frac{1}{P^{(+)}_{I_{k,q}}\,(\sum_{i \subset I_{k,q} }H_i-z)\,P^{(+)}_{I_{k,q}}})^{\frac{1}{2}}\Big\}^{l}\nonumber \\
& &\times (\frac{1}{P^{(+)}_{I_{k,q}}\,(\sum_{i\subset I_{k,q} }H_i-z)\,P^{(+)}_{I_{k,q}}})^{\frac{1}{2}}\,.\label{exp-G-fin}
\end{eqnarray}
To justify this, we need  some ingredients. First,  we make use of the spectral theorem (recall that $H_{I_{r,i}}^0$ is self-adjoint) and the assumption in (\ref{gaps}) to derive the bound
\begin{equation}\label{spectral-th}
\|P^{(+)}_{I_{r,i}}\,(\frac{H_{I_{r,i}}^0+1}{H_{I_{r,i}}^0})^{\frac{1}{2}}P^{(+)}_{I_{r,i}}\|\leq \sqrt{2}\,.
\end{equation}
By combining (\ref{ass-2}) and (\ref{ass-2-bis}), we then find that
\begin{equation}
\|(H_{I_{j,i}}^0+1)^{-\frac{1}{2}}(V^{(k,q-1)}_{I_{j, i}}-\langle V^{(k,q-1)}_{I_{j,i}} \rangle)(H_{I_{j,i}}^0+1)^{-\frac{1}{2}}\|\leq 2 |\tau|^{\frac{j-1}{4}}\,.\label{est-corr}
\end{equation}
Next, for each $1\leq j \leq k-1$, we can make use of the following inequality
\begin{eqnarray}
&& \|(\frac{1}{P^{(+)}_{I_{k,q}}\,(\sum_{i \subset I_{k,q} }H_i-z)\,P^{(+)}_{I_{k,q}}})^{\frac{1}{2}}\,P^{(+)}_{I_{k,q}}[P^{(+)}_{I_{j,i}}(V^{(k,q-1)}_{I_{j, i}}-\langle V^{(k,q-1)}_{I_{j,i}} \rangle) P^{(+)}_{I_{j,i}}]P^{(+)}_{I_{k,q}}\,(\frac{1}{P^{(+)}_{I_{k,q}}\,(\sum_{i \subset I_{k,q} }H_i-z)\,P^{(+)}_{I_{k,q}}})^{\frac{1}{2}}\| \nonumber \\
&\leq &\|(\frac{1}{P^{(+)}_{I_{k,q}}\,(\sum_{i \subset I_{k,q} }H_i-z)\,P^{(+)}_{I_{k,q}}})^{\frac{1}{2}}\,P^{(+)}_{I_{k,q}}P^{(+)}_{I_{j , i}}\,(H_{I_{j,i}}^0)^{\frac{1}{2}}(\frac{H_{I_{j,i}}^0+1}{H_{I_{j,i}}^0})^{\frac{1}{2}}\|^2\times\\
& &\quad \times \|(H_{I_{j,i}}^0+1)^{-\frac{1}{2}}(V^{(k,q-1)}_{I_{j, i}}-\langle V^{(k,q-1)}_{I_{j,i}} \rangle)(H_{I_{j,i}}^0+1)^{-\frac{1}{2}}\|\\
&\leq&\|(\frac{1}{P^{(+)}_{I_{k,q}}\,(\sum_{i \subset I_{k,q} }H_i-z)\,P^{(+)}_{I_{k,q}}})^{\frac{1}{2}}\,P^{(+)}_{I_{k,q}}(H_{I_{j,i}}^0)^{\frac{1}{2}}\|^2\,\|P^{(+)}_{I_{j,i}}\,(\frac{H_{I_{j,i}}^0+1}{H_{I_{j,i}}^0})^{\frac{1}{2}}\|^2\times\\
& &\quad \times \|(H_{I_{j,i}}^0+1)^{-\frac{1}{2}}(V^{(k,q-1)}_{I_{j, i}}-\langle V^{(k,q-1)}_{I_{j,i}} \rangle)(H_{I_{j,i}}^0+1)^{-\frac{1}{2}}\|\\
&=&\|(\frac{1}{P^{(+)}_{I_{k,q}}\,(\sum_{i \subset I_{k,q} }H_i-z)\,P^{(+)}_{I_{k,q}}})^{\frac{1}{2}}\,P^{(+)}_{I_{k,q}}H_{I_{j,i}}^0\,P^{(+)}_{I_{k,q}}(\frac{1}{P^{(+)}_{I_{k,q}}\,(\sum_{i \subset I_{k,q} }H_i-z)\,P^{(+)}_{I_{k,q}}})^{\frac{1}{2}}\|\times\\
& &\,\times \|P^{(+)}_{I_{j,i}}\,(\frac{H_{I_{j,i}}^0+1}{H_{I_{j,i}}^0})^{\frac{1}{2}}\|^2\,\|(H_{I_{j,i}}^0+1)^{-\frac{1}{2}}(V^{(k,q-1)}_{I_{j, i}}-\langle V^{(k,q-1)}_{I_{j,i}} \rangle)(H_{I_{j,i}}^0+1)^{-\frac{1}{2}}\|\,.
\end{eqnarray}
Recalling (\ref{spectral-th}) and (\ref{est-corr}), we finally derive the bound
\begin{eqnarray}
& &\|(\frac{1}{P^{(+)}_{I_{k,q}}\,(\sum_{i \subset I_{k,q} }H_i-z)\,P^{(+)}_{I_{k,q}}})^{\frac{1}{2}}\,P^{(+)}_{I_{k,q}}[P^{(+)}_{I_{j,i}}(V^{(k,q-1)}_{I_{j, i}}-\langle V^{(k,q-1)}_{I_{j,i}} \rangle) P^{(+)}_{I_{j,i}}]P^{(+)}_{I_{k,q}}\,(\frac{1}{P^{(+)}_{I_{k,q}}\,(\sum_{i \subset I_{k,q} }H_i-z)\,P^{(+)}_{I_{k,q}}})^{\frac{1}{2}}\|\nonumber \\
&\leq&4\,|\tau|^{\frac{j-1}{4}}\,\|(\frac{1}{P^{(+)}_{I_{k,q}}\,(\sum_{i \subset I_{k,q} }H_i-z)\,P^{(+)}_{I_{k,q}}})^{\frac{1}{2}}\,P^{(+)}_{I_{k,q}}H_{I_{j,i}}^0\,P^{(+)}_{I_{k,q}}(\frac{1}{P^{(+)}_{I_{k,q}}\,(\sum_{i \subset I_{k,q} }H_i-z)\,P^{(+)}_{I_{k,q}}})^{\frac{1}{2}}\|\,.\label{intermediate}
\end{eqnarray}
We observe that,  for $1\leq l \leq L \leq N-r$, 
\begin{equation}\label{ineq-inter-00}
\sum_{i=l}^{L}H^{0}_{I_{r,i}}\leq (r+1) \sum_{i=l}^{L+r} H_i \,,
\end{equation}
and,  for $|z|\leq \frac{1}{2}$,
\begin{equation}\label{intermediate-sum}
\|(\frac{1}{P^{(+)}_{I_{k,q}}\,(\sum_{i \subset I_{k,q} }H_i-z)\,P^{(+)}_{I_{k,q}}})^{\frac{1}{2}}\,P^{(+)}_{I_{k,q}}\,(\sum_{i \subset I_{k,q} }H_i)\,P^{(+)}_{I_{k,q}}(\frac{1}{P^{(+)}_{I_{k,q}}\,(\sum_{i \subset I_{k,q} }H_i-z)\,P^{(+)}_{I_{k,q}}})^{\frac{1}{2}}\|\leq 2\,,
\end{equation}
which follows from  the spectral theorem and from the assumption in (\ref{gaps}).
Hence,  it readily follows from (\ref{intermediate}) that
\begin{eqnarray}
& &\|\sum_{I_{j,i} \subset I_{k,q}}(\frac{1}{P^{(+)}_{I_{k,q}}\,(\sum_{i \subset I_{k,q} }H_i-z)\,P^{(+)}_{I_{k,q}}})^{\frac{1}{2}}\,P^{(+)}_{I_{k,q}}[P^{(+)}_{I_{j,i}}(V^{(k,q-1)}_{I_{j, i}}-\langle V^{(k,q-1)}_{I_{j,i}} \rangle) P^{(+)}_{I_{j,i}}]P^{(+)}_{I_{k,q}}\,(\frac{1}{P^{(+)}_{I_{k,q}}\,(\sum_{i \subset I_{k,q} }H_i-z)\,P^{(+)}_{I_{k,q}}})^{\frac{1}{2}}\|\nonumber \\
&\leq&4\,\sum_{j=1}^{k-1}|\tau|^{\frac{j-1}{4}}\,\|(\frac{1}{P^{(+)}_{I_{k,q}}\,(\sum_{i \subset I_{k,q} }H_i-z)\,P^{(+)}_{I_{k,q}}})^{\frac{1}{2}}\,P^{(+)}_{I_{k,q}}\sum_{I_{j,i} \subset I_{k,q}}H_{I_{j,i}}^0\,P^{(+)}_{I_{k,q}}(\frac{1}{P^{(+)}_{I_{k,q}}\,(\sum_{i \subset I_{k,q} }H_i-z)\,P^{(+)}_{I_{k,q}}})^{\frac{1}{2}}\|\,\\
&\leq &4\,\sum_{j=1}^{k-1}|\tau|^{\frac{j-1}{4}}(j+1)\,\|(\frac{1}{P^{(+)}_{I_{k,q}}\,(\sum_{i \subset I_{k,q} }H_i-z)\,P^{(+)}_{I_{k,q}}})^{\frac{1}{2}}\,P^{(+)}_{I_{k,q}}\,(\sum_{i \subset I_{k,q} }H_i)\,P^{(+)}_{I_{k,q}}(\frac{1}{P^{(+)}_{I_{k,q}}\,(\sum_{i \subset I_{k,q} }H_i-z)\,P^{(+)}_{I_{k,q}}})^{\frac{1}{2}}\|\\
&\leq &8\,\sum_{j=1}^{+\infty}(j+1)\,|\tau|^{\frac{j-1}{4}}\,,
\end{eqnarray}
where we have used (\ref{intermediate-sum}) in the last step.
Finally,  we can conclude that, for $|\tau|$ sufficiently small but independent of $k,q$, and $N$,
\begin{eqnarray}
& &\|\sum_{l=0}^{\infty}\,\Big\{(\frac{1}{P^{(+)}_{I_{k,q}}\,(\sum_{i \subset I_{k,q} }H_i-z)\,P^{(+)}_{I_{k,q}}})^{\frac{1}{2}}\,[P^{(+)}_{I_{k, q}}\,\tau \mathcal{V}_{I_{k,q}}\,P^{(+)}_{I_{k,q}}]\,(\frac{1}{P^{(+)}_{I_{k,q}}\,(\sum_{i \subset I_{k,q} }H_i-z)\,P^{(+)}_{I_{k,q}}})^{\frac{1}{2}}\Big\}^{l}\|\quad\quad \label{est-exp-in}\quad \\
&\leq&\frac{1}{1-8|\tau|\,\sum_{j=1}^{+\infty}(j+1)\,|\tau|^{\frac{j-1}{4}}}<\infty\,.\label{est-exp-fin}
\end{eqnarray}

\qed

Lemma \ref{gap} implies that, under  assumption (\ref{ass-2}), $E_{I_{k,q}}$ is an eigenvalue of the Hamiltonian $G_{I_{k,q}}=G_{I_{k,q}}(\tau)$ isolated from the rest of its spectrum by a distance larger than or equal to $\frac{1}{2}$,  for $|\tau|$ sufficiently small but \textit{independent} of $N$,  $k$, and $q$; as  stated in the following Corollary.
\begin{cor}\label{cor-gap}
Assuming (\ref{ass-2}), and choosing $|\tau|$ sufficiently small, but independent of $N$, $k$,  and $q$,  the following statement holds: the spectrum of the Hamiltonian $G_{I_{k,q}}$  in the disk of radius $\frac{1}{2}$ centred at $E_{I_{k,q}}$ consists of only $E_{I_{k,q}}$,  and $E_{I_{k,q}}$ is the eigenvalue of $G_{I_{k,q}}$ corresponding to the ``vacuum'' eigenvector, $\bigotimes_{j\in I_{k,q}}\Omega_{j}$\,\,, in $\mathcal{H}_{I_{k,q}}$, i.e.,
\begin{eqnarray}
G_{I_{k,q}}P^{(-)}_{I_{k,q}}
&= &E_{I_{k,q}}P^{(-)}_{I_{k,q}}\nonumber\\
&= &\, \Big[\tau\sum_{ I_{1,i}\subset I_{k,q}}\langle  V^{(k,q-1)}_{I_{1,i}} \rangle+\dots+\tau\sum_{I_{k-1,i}\subset  I_{k,q}}\langle V^{(k,q-1)}_{I_{k-1,i}}\rangle \Big]P^{(-)}_{I_{k,q}}\,,\label{final-eq-2}
\end{eqnarray}
and
\begin{eqnarray}
G_{I_{k,q}}P^{(+)}_{I_{k,q}}
&= &P^{(+)}_{I_{k,q}}G_{I_{k,q}}P^{(+)}_{I_{k,q}}\nonumber
\end{eqnarray}
with $P^{(+)}_{I_{k,q}}(G_{I_{k,q}}-E_{I_{k,q}}-z)P^{(+)}_{I_{k,q}}$ invertible on $P^{(+)}_{I_{k,q}}\mathcal{H}$ for $|z|\leq \frac{1}{2}$.
\end{cor}

\subsection{Block-diagonalization: Bound on $\|V^{(k,q)}_{I_{r,i}}\|_{H^0}$ and consistency of the iterative scheme }\label{block-bound}

%In the next theorem, we estimate the weighted norm $$\|V^{(k,q)}_{I_{r,i}}\|_{H^0}:=\|(H_{I_{r,i}}^0+1)^{-\frac{1}{2}}V^{(k,q)}_{I_{r,i}}(H_{I_{r,i}}^0+1)^{-\frac{1}{2}}\|$$ in terms of $\|V^{(k,q-1)}_{I_{r,i}}\|_{H^0}$. For a fixed interval $I_{r,i}$, the weighted norm of the potential does not change, i.e., $\|V^{(k,q-1)}_{I_{r,i}}\|_{H^0}=\|V^{(k,q)}_{I_{r,i}}\|_{H^0}$, 
%in the step $(k,q-1) \rightarrow (k,q)$, unless some conditions are fulfilled. To gain some intuition of this fact, the reader is advised to take a look at Fig. 1, (replacing $l$ by $r$). Notice that shifting the interval $I_{k,q}$ to the left by one site makes it coincide with $I_{k,q-1}$. If $I_{k,q}$ is not contained in $I_{r,i}$   then  $\|V^{(k,q)}_{I_{r,i}}\|_{H^0}=\|V^{(k,q-1)}_{I_{r,i}}\|_{H^0}$. Therefore, in the step $(k,q-1) \rightarrow (k,q)$,  a change of the weighted norm, i.e., 
%$\|V^{(k,q)}_{I_{r,i}}\|_{H^0}\neq \|V^{(k,q-1)}_{I_{r,i}}\|_{H^0}$, may only happen in  at most $r-k+1$ cases, provided $r> k$, and only in one case if $k$ coincides with the length $r$\,; and it never happens if $r<k$.  

Recall that, according to the rules of the algorithm, the weighted norm of the potentials does not change, i.e., $\|V^{(k,q-1)}_{I_{r,i}}\|_{H^0}=\|V^{(k,q)}_{I_{r,i}}\|_{H^0}$, 
in the step $(k,q-1) \rightarrow (k,q)$, unless $I_{r,i}\cap I_{k,q}\neq \emptyset$; for more details see  \cite{DFPR}.

\noindent
In the theorem below we estimate the change of the norm of the potentials in the block-diagonalization steps, for each $k$, starting from $k=0$.  We have to make use of a lower bound on the distance between $E_{I_{k,q}}$ and the rest of the spectrum  of the operator $G_{I_{k,q}}$. 
%This lower bound follows from estimate (\ref{ass-2}), as explained in Lemma \ref{gap} and Corollary \ref{cor-gap}. 
We will proceed inductively by showing that, for $|\tau|$ sufficiently small but independent of $r$, $N$, $k$, and $q$, the operator-norm bound in (\ref{ass-2}), at step $(k,q-1)$, $q\geq 2$  (for $q=1$ see the footnote), yields control over the spectrum of the Hamiltonian $G_{I_{k,q}}$ in a disk centred at $E_{I_{k,q}}$, (see Corollary \ref{cor-gap}), and the latter provides an essential  ingredient for the proof of a bound on the weighted  operator norms of the potentials, according to  (\ref{ass-2}), at the next step\footnote{  \label{footnote} Recall the special steps of type $(k-1, N-k+1) \rightarrow (k,1)$.} $(k,q)$.\\

\begin{thm}\label{norms}
 Assume that $|\tau|\leq t_0$, with $t_0$ sufficiently small but independent of $k$, $q$, and $N$ and such that the assumptions of Lemma \ref{unboundedlemmaA3} are fulfilled. Then the Hamiltonians $G_{I_{k,q}}$ are well-defined closed operators, and
\begin{enumerate}
\item[S1)] for any interval  $I_{r,i}$, with $r\geq 1$, for $(k,q)\prec (r,i+1)$ and for $(k,q)=(r,i+1)$ the operator $$(H_{I_{r,i}}^0+1)^{-\frac{1}{2}}V^{(k,q)}_{I_{r,i}}(H_{I_{r,i}}^0+1)^{-\frac{1}{2}}$$ has a norm bounded by $|\tau|^{\frac{r-1}{4}}$,
 \item[S2)]$G_{I_{k,q+1}}P^{(+)}_{I_{k,q+1}}$ has no spectrum in the disk $\overline{\mathbb{D}_{\frac{1}{2}}}$ centered at $E_{I_{k,q+1}}$ where  $G_{I_{k,q}}$ is defined in (\ref{def-G}) for $k\geq 2$,  and $G_{I_{1,q}}:=H_{q}+H_{q+1}$.
\end{enumerate}
\end{thm}

%\newpage
\noindent
\emph{Proof.}

\noindent
The proof is identical to Theorem 4.1 in \cite{DFPR}, provided $t$ is replaced  by $\tau$ or  by $|\tau|$, respectively, depending on the context.
\qed

In the next theorem we explain how the Hamiltonian $K_{N}^{(k,q)}$  (see (\ref{def-transf-ham})) is defined  in terms of the potentials $V_{I_{l,j}}^{(k,q)}$ (see Definition \ref{def-interections})  and prove that, as an operator, it coincides with $e^{S_{I_{k,q}}}\,K_N^{(k,q-1)}\,e^{-S_{I_{k,q}}}$.

\begin{thm}\label{th-potentials}
Assume that $|\tau|\leq t_0$. For the Hamiltonian $K_{N}^{(k,q)}$ (see (\ref{def-transf-ham})), with $k\geq1$ and $q\geq 2$, the following identity
 \begin{eqnarray}
e^{S_{I_{k,q}}}\,K_N^{(k,q-1)}\,e^{-S_{I_{k,q}}}
& =&\sum_{i=1}^{N}H_{i}+\tau \sum_{i=1}^{N-1}V^{(k,q)}_{I_{1,i}}+\tau\sum_{i=1}^{N-2}V^{(k,q)}_{I_{2,i}}+\dots+\tau\sum_{i=1}^{N-k}V^{(k,q)}_{I_{k,i}} \\
& &+\tau\sum_{i=1}^{N-k-1}V^{(k,q)}_{I_{k+1,i}}+\dots+\tau\sum_{i=1}^{2}V^{(k,q)}_{I_{N-2,i}}+\tau V^{(k,q)}_{I_{N-1,1}}\label{def-transf-ham-bis}
\end{eqnarray}
holds on the domain $e^{S_{I_{k,q}}}D(K_{N}^{(k,q-1)})$, where the operator
on the r-h-s is understood as the unique $m$-sectorial operator associated with the $m$-sectorial form 
$$\kappa_{N}^{(k,q)}: D((H^0_{I_{N-1,1}})^{\frac{1}{2}})\times D((H^0_{I_{N-1,1}})^{\frac{1}{2}})\rightarrow\mathbb{C}$$ given by
\begin{eqnarray*}
\kappa_{N}^{(k,q)}(\varphi,\psi):=&&\sum_{i=1}^{N}\langle H_{i}^{\frac{1}{2}}\varphi, H_{i}^{\frac{1}{2}}\psi \rangle +\tau \sum_{i=1}^{N-1}\langle V^{(k,q)}_{I_{1,i}}\varphi,\psi  \rangle+\tau\sum_{i=1}^{N-2}\langle V^{(k,q)}_{I_{2,i}}\varphi, \psi  \rangle +\dots+\tau\sum_{i=1}^{N-k}\langle V^{(k,q)}_{I_{k,i}}\varphi, \psi  \rangle \\
&&+\tau\sum_{i=1}^{N-k-1} \langle V^{(k,q)}_{I_{k+1,i}}\varphi, \psi  \rangle+\dots+\tau\sum_{i=1}^{2} \langle V^{(k,q)}_{I_{N-2,i}}\varphi, \psi  \rangle+\tau \langle V^{(k,q)}_{I_{N-1,1}}\varphi, \psi  \rangle\,
\end{eqnarray*}
with $ \varphi, \psi\in D((H^0_{I_{N-1,1}})^{\frac{1}{2}})$.

%where the operators $\{V^{(k,q)}_{I_{l,i}}\}$ are determined by the operators $\{V^{(k,q-1)}_{I_{l,i}}\}$ of the previous iteration step, as specified in  Definition \ref{def-interections}. If $q=1$ the statement holds with $(k,q-1)$ replaced by $(k-1, N-k+1)$. 
\end{thm}

%\newpage
\noindent
\emph{Proof.}

\noindent
We present the proof for $q\geq 2$, the case $q=1$ can be proved in the same way. We observe that, by following the arguments used in Theorem 4.2 of  \cite{DFPR}, one can prove that the identity claimed in the statement holds formally. However, our final goal is to prove that (\ref{def-transf-ham-bis}) is in fact an identity between two $m$-sectorial operators (for definitions and results on $m$-sectorial operators the reader is referred to the classic monography by T. Kato \cite{K}). \\ 
To this aim, we first observe that $D((H^0_{I_{N-1,1}})^{\frac{1}{2}})$ is invariant under the action of $e^{S_{I_{k,q}}}$. This is shown by the following estimate.\\  

For any $\varphi \in D((H^0_{I_{N-1,1}})^{\frac{1}{2}})$ and $m\in \mathbb{N}$ we have
\begin{eqnarray}
& &\|(H^0_{I_{r,i}})^{\frac{1}{2}}\,(S_{I_{k;q}})^m\varphi\| \label{domain-in}\\
&=&\|(H^0_{I_{r,i}})^{\frac{1}{2}}\,\frac{1}{(H^0_{I_{r,i}\setminus I_{k,q}}+1)^{\frac{1}{2}}}(H^0_{I_{r,i}\setminus I_{k,q}}+1)^{\frac{1}{2}}(S_{I_{k;q}})^m\varphi\| \\
&=&\|(H^0_{I_{r,i}})^{\frac{1}{2}}\,\frac{1}{(H^0_{I_{r,i}\setminus I_{k,q}}+1)^{\frac{1}{2}}}  (S_{I_{k;q}})^m \,(H^0_{I_{r,i}\setminus I_{k,q}}+1)^{\frac{1}{2}}\varphi\|\\
&\leq &\|\frac{(H^0_{I_{r,i}})^{\frac{1}{2}}}{(H^0_{I_{r,i}\setminus I_{k,q}}+1)^{\frac{1}{2}}(H^0_{ I_{k,q}}+1)^{\frac{1}{2}}}\|\,\|(H^0_{I_{k,q}}+1)^{\frac{1}{2}}S_{I_{k;q}}\|\,\|S_{I_{k;q}}\|^{m-1}\,\|(H^0_{I_{r,i}\setminus I_{k,q}}+1)^{\frac{1}{2}}\varphi \| \\
&\leq &C_{\varphi}^m\,, \label{domain-fin}
\end{eqnarray}
for some constant $C_{\varphi}$ depending on $\varphi$. Here we have exploited  estimates (\ref{bound-S}) and (\ref{S-Hest}) in Lemma \ref{unboundedlemmaA3}, the spectral theorem for commuting self-adjoint operators, and the assumption $\varphi \in D((H^0_{I_{N-1,1}})^{\frac{1}{2}})$.

%{\color{blue}Using the results and the arguments of Theorem \ref{th-norms}, the identities in (\ref{local-ham}) and (\ref{growth-bis}) hold as operator identities if we sandwich both the l-h-s and the r-s-d with $(H_{I_{k,q}}^0+1)^{-\frac{1}{2}}$ and $(H_{I_{l,i}}^0+1)^{-\frac{1}{2}}$, respectively, i.e., 
%\begin{equation}
%(H_{I_{l,i}}^0+1)^{-\frac{1}{2}}e^{S_{I_{k;q}}}\,V^{(k,q-1)}_{I_{l;i}}\,e^{-S_{I_{k;q}}}(H_{I_{l,i}}^0+1)^{-\frac{1}{2}}=(H_{I_{l,i}}^0+1)^{-\frac{1}{2}}\Big\{V^{(k,q-1)}_{I_{l;i}}+\sum_{n=1}^{\infty}\frac{1}{n!}\,ad^{n}S_{I_{k;q}}(V^{(k,q-1)}_{I_{l;i}})\Big\}(H_{I_{l,i}}^0+1)^{-\frac{1}{2}} \,.\label{growth-bis-bis}
%\end{equation}
Next, using the type of manipulations and estimates appearing in the proof of Theorem \ref{norms},  we derive that the relation in (\ref{def-transf-ham-bis})  holds as an identity between matrix elements with vectors in the domain $D((H^0_{I_{N-1, 1}})^{\frac{1}{2}})$, i.e., on the l-h-s of (\ref{def-transf-ham-bis}) we can expand the exponential operator and control the series whenever we consider a matrix element with vectors $\varphi $, $\psi $ in $D((H^0_{I_{N-1, 1}})^{\frac{1}{2}})$, and then check that they correspond to the analogous matrix elements of the terms on the r-h-s. In this step one has to make sure that  also  the off-diagonal terms  that cancel out on the l-h-s of (\ref{def-transf-ham-bis}) are individually  well defined; (this cancellation is indeed the purpose of the conjugation).

Now, thanks to the estimate provided by S1) in Theorem \ref{norms} and \cite[Theorem 3.9, p. 340]{K}, the bilinear form $\kappa_{N}^{(k,q)}$ defined in the statement 
 is $m$-sectorial on its domain as a perturbation of a closed symmetric form  (the one determined by
$H^0_{I_{N-1,1}}$) by a small perturbation in the sense of quadratic forms,  provided $|\tau|$ is chosen small enough, but independent of either $N$ or $(k,q)$.
Next, we invoke \cite[Theorem 2.1, p. 322] {K} to define $K_N^{(k,q)}$ as the unique $m$-sectorial operator associated with the form $\kappa_{N}^{(k,q)}$ whose domain,
$D(K_N^{(k,q)})$,  is contained in the domain of the form itself, namely in $D((H^0_{I_{N-1,1}})^{\frac{1}{2}})$.
Using an induction,  we see that the operator $e^{S_{I_{k;q}}}\,K_N^{(k,q-1)}\,e^{-S_{I_{k;q}}}$ is  $m$-sectorial, since $K_N$ is m-sectorial. In light of the invariance property (see (\ref{domain-in})-(\ref{domain-fin})) proved above, its
domain is contained in $D((H^0_{I_{N-1,1}})^{\frac{1}{2}})$.  The identity between matrix elements discussed above implies that the form determined by $e^{S_{I_{k;q}}}\,K_N^{(k,q-1)}\,e^{-S_{I_{k;q}}}$ coincides with the restriction to $e^{S_{I_{k;q}}}\,D(K_N^{(k,q-1)})$  of the form determined by $K_N^{(k,q)}$. Hence, a straightforward application of \cite[Corollary 2.4, p. 323]{K} shows that  the operator  $e^{S_{I_{k;q}}}\,K_N^{(k,q-1)}\,e^{-S_{I_{k;q}}}$ is extended by $K_N^{(k,q)}$. By recalling that $m$-sectorial operators are  $m$-accretive (cf. p. 279-280 in \cite{K}), and no proper inclusions can hold between any  two $m$-accretive operators, we conclude that (\ref{def-transf-ham-bis}) is  an identity between two $m$-sectorial operators. We recall that $m$-sectorial operators are densely defined; in particular from \cite[Corollary 2.4, p. 323]{K} we get that $D(H^0_{I_{N-1,1}})\subseteq D(K_N^{(k,q)})$, using induction,  since $K_N^{(0,N)}\equiv 
 K_N$ and $D(H^0_{I_{N-1,1}})\subseteq D(K_N)$, as explained in Section  \ref{intro-def}.

\qed

\begin{thm}\label{main-res}
Under the assumption that (\ref{gaps}), (\ref{potential}) and (\ref{klmn-cond}) hold, the complex Hamiltonian $K_{N}\equiv K_N(\tau)$ defined in (\ref{complex-Hamiltonian}) has the following properties: There exists some $t_0 > 0$ such that, for any $\tau\in \mathbb{C}$ with 
$|\tau | \leq t_0$, and for all $N < \infty$, an invertible operator $U_N(\tau)$ can be constructed such that
\begin{enumerate}
\item $U^{-1}_N(\tau)K_N(\tau)U_N(\tau)$ has a nondegenerate eigenvalue, $E_N(\tau)$;
\item the rest of its spectrum is at a distance  larger than or equal to $\frac{1}{2}$ from $E_N(\tau)$;
\item for $\tau \equiv t\in \mathbb{R}$, the eigenvalue $E_N(\tau\equiv t)$ is the nondegenerate ground-state energy of $K_{N}(t)$.
\end{enumerate}

\end{thm}

\noindent
\emph{Proof.}
Notice that $K_N^{(N-1,1)} \equiv G_{I_{N-1,1}}+\tau V^{(N-1,1)}_{I_{N-1,1}}$. Thus, we have constructed the invertible operator 
$U_{N}(\tau)$,  see (\ref{conjug}), such that  the operator
$$U^{-1}_{N}(\tau)K_{N}(\tau)U_{N}(\tau)=G_{I_{N-1,1}}+\tau V^{(N-1,1)}_{I_{N-1,1}}=: \widetilde{K}_{N}(\tau)$$  
has the properties in (\ref{block-diag-eq}) and (\ref{gapss}), which follow from Theorem \ref{norms} and from (\ref{final-eq-1}) and (\ref{final-eq-2}), for $(k,q)=(N-1,1)$, where we also include the block-diagonalized potential $V^{(N-1,1)}_{I_{N-1,1}}$. \qed 
\begin{rem}\label{comment-t0}
We stress that, for $\tau$ in the disk of radius $t_0$ centered at $0$, all the series used in the construction of all the intermediate Hamiltonians converge uniformly. This result will be invoked in the next theorem wherever those series appear.
\end{rem}
\subsection{Block-diagonalization: Analyticity of $E_N\equiv E_N(\tau)$}\label{block-analyticity}

\begin{thm}\label{analyticity}
Under the hypotheses of Theorem \ref{main-res}, the eigenvalue $E_N(\tau )$ of $U^{-1}_{N}(\tau)K_N(\tau)U_{N}(\tau)$ is an analytic function of $\tau $ in $\mathbb{D}_{t_0}:=\{\tau \in \mathbb{C}\,:\, |\tau|<t_0\}$.
\end{thm}

\noindent
\emph{Proof}

\noindent
Since, by construction, $E_N(\tau):= \langle \widetilde K_N(\tau)\Omega, \Omega \rangle$, with 
\begin{eqnarray}
\widetilde K_N(\tau):=K_N^{(N-1,1)}
& :=&\sum_{i=1}^{N}H_{i}+\tau\sum_{i=1}^{N-1}V^{(N-1,1)}_{I_{1,i}}+\tau\sum_{i=1}^{N-2}V^{(N-1,1)}_{I_{2,i}}+\dots+\tau V^{(N-1,1)}_{I_{N-1,1}}\,,
\end{eqnarray}
it is enough to show that, for all $1\leq i \leq N-r$, $1\leq r \leq N-1$,  the operator-valued functions
\begin{equation}
(\frac{1}{H^0_{I_{r,i}}+1})^{\frac{1}{2}}V_{I_{r,i}}^{(N-1,1)}(\frac{1}{H^0_{I_{r,i}}+1})^{\frac{1}{2}}
\end{equation}
are analytic in $\mathbb{D}_{t_0}$.
Our strategy to show this will consist in establishing the following property: 
\\

\noindent
{\bf{Property A}}
\emph{ For any $I_{r,i}$ and $(k,q)$,
the (bounded) operators 
\begin{equation}\label{analticity}
(\frac{1}{H^0_{I_{r,i}}+1})^{\frac{1}{2}}V_{I_{r,i}}^{(k,q)}(\frac{1}{H^0_{I_{r,i}}+1})^{\frac{1}{2}}\,\equiv \, (\frac{1}{H^0_{I_{r,i}}+1})^{\frac{1}{2}}V_{I_{r,i}}^{(k,q)}(\tau)(\frac{1}{H^0_{I_{r,i}}+1})^{\frac{1}{2}}
\end{equation}
%\item[ii)] The (bounded) operators $S_{I_{r,i}}\equiv S_{I_{r,i}}(\tau) $
 %are  (strongly) analytic operator-valued functions;
%\item[iii)]  The operator valued functions
%\begin{equation}
%\frac{1}{G_{I_{r,i}}-E_{I_{r,i}}}P^{(+)}_{I_{r,i}}
%\end{equation}  are analytic.
%\end{enumerate}
are (strongly) analytic in $\tau \in \mathbb{D}_{t_0}$.}
\\

\noindent
The result will be achieved by implementing an inductive argument in $(k, q)$. Since we will deal with bounded operators, the analyticity is always understood in the strong sense. \\

\noindent
\emph{Initial step}\\

At the initial step, $(0,N)$, the only nonzero potentials are $V_{I_{1,j}}^{(0, N)}$, $1\leq j\leq N-1$, and they are $\tau$-independent. 
Hence the corresponding operators $$(\frac{1}{H^0_{I_{1,j}}+1})^{\frac{1}{2}}V_{I_{1,j}}^{(0, N)}(\frac{1}{H^0_{I_{1,j}}+1})^{\frac{1}{2}}$$ are entire operator-valued functions.\\

\noindent
\emph{Inductive step}\\

We can now move on to the inductive step and show that Property A holds  for all $ I_{r,i}$ at step $(k,q)$ if it holds for all $(k', q')\prec  (k,q)$. There are four different cases, a), b), c), and d-1), d-2), as in Definition 
\ref{def-interections}. We observe that case a) is trivial.  Next, we study  case b) in detail. 
\\

\noindent
\emph{Case b)}

\noindent
We want to prove that 
\begin{equation}
(\frac{1}{H^0_{I_{k,q}}+1})^{\frac{1}{2}}V^{(k,q)}_{I_{k,q}}(\frac{1}{H^0_{I_{k,q}}+1})^{\frac{1}{2}}
\end{equation}
is analytic in $\tau\in \mathbb{D}_{t_0}$. We  recall the formula
\begin{equation}
V_{k,q}^{(k,q)}=\sum_{j=1}^\infty \tau^{j-1}(V_{k,q}^{(k,q-1)})_j^{diag}\,,
\end{equation}
where the terms
$(V_{k,q}^{(k,q-1)})_j$ are given by
\begin{eqnarray}\label{Vj-bis}
& &(V^{(k, q-1)}_{I_{k,q}})_j\,:=\label{formula-v_j-bis}\\
& &\sum_{p\geq 2, r_1\geq 1 \dots, r_p\geq 1\, ; \, r_1+\dots+r_p=j}\frac{1}{p!}\text{ad}\,(S_{I_{k,q}})_{r_1}\Big(\text{ad}\,(S_{I_{k,q}})_{r_2}\dots (\text{ad}\,(S_{I_{k,q}})_{r_p}(G_{I_{k,q}}) \Big)\\
& &+\sum_{p\geq 1, r_1\geq 1 \dots, r_p\geq 1\, ; \, r_1+\dots+r_p=j-1}\frac{1}{p!}\text{ad}\,(S_{I_{k,q}})_{r_1}\Big(\text{ad}\,(S_{I_{k,q}})_{r_2}\dots (\text{ad}\,(S_{I_{k,q}})_{r_p}(V^{(k,q-1)}_{I_{k,q}}) \Big)\,,\quad\quad\quad\quad\,.
\end{eqnarray}
and the bounded 
operators $(S_{I_{k,q}})_j$ are computed by means of the following formula
\begin{equation}\label{S_j}
(S_{I_{k,q}})_j:=\frac{1}{G_{I_{k,q}}-E_{I_{k,q}}}P^{(+)}_{I_{k,q}}\,(V^{(k, q-1)}_{I_{k,q}})_j\,P^{(-)}_{I_{k,q}}-P^{(-)}_{I_{k,q}}(V^{(k, q-1)}_{I_{k,q}})_j\frac{1}{G_{I_{k,q}}-E_{I_{k,q}}}P^{(+)}_{I_{k,q}}\,.
\end{equation}
Similarly to the control of $(V^{(k, q-1)}_{I_{k,q}})_j$ we insert the identity operator in the form
\begin{equation}\label{V-J-sandwich}
\charf=(\frac{1}{H^0_{I_{k,q}}+1})^{\frac{1}{2}}(H^0_{I_{k,q}}+1)^{\frac{1}{2}}=(H^0_{I_{k,q}}+1)^{\frac{1}{2}}(\frac{1}{H^0_{I_{k,q}}+1})^{\frac{1}{2}}
\end{equation}
in a suitable way to express (\ref{Vj-bis}) in terms of the operators
\begin{equation}\label{V-J-sandwich}
(\frac{1}{H^0_{I_{k,q}}+1})^{\frac{1}{2}}V^{(k, q-1)}_{I_{k,q}}(\frac{1}{H^0_{I_{k,q}}+1})^{\frac{1}{2}}
\end{equation}
and
\begin{equation}
(S_{I_{k,q}})_r(H^0_{I_{k,q}}+1)^{\frac{1}{2}}\,,\,(H^0_{I_{k,q}}+1)^{\frac{1}{2}}(S_{I_{k,q}})_r
\end{equation}
with $r<j$.
Then, assuming that Property A holds at step $(k,q-1)$, saying that, for all $I_{r,i}$ and for all $(k',q') \prec (k,q)$, the operators
\begin{equation}
(\frac{1}{H^0_{I_{r,i}}+1})^{\frac{1}{2}}V^{(k, q-1)}_{I_{r,i}}(\frac{1}{H^0_{I_{r,i}}+1})^{\frac{1}{2}}
\end{equation}
are analytic in $ \mathbb{D}_{t_0}$,   we can implement an induction on $j$ and prove that, for all $j$, the operators 
\begin{equation}\label{V-J-sandwich}
(\frac{1}{H^0_{I_{k,q}}+1})^{\frac{1}{2}}(V^{(k, q-1)}_{I_{k,q}})_j(\frac{1}{H^0_{I_{k,q}}+1})^{\frac{1}{2}}
\end{equation} are analytic in  $\tau\in \mathbb{D}_{t_0}$, where the result for $j=1$ is precisely Property A at step $(k,q-1)$. Indeed, suppose it is true for all $j'<j$, then, using Property A at step $(k,q-1)$, we get that the operators
\begin{equation}
(S_{I_{k,q}})_{j'}(H^0_{I_{k,q}}+1)^{\frac{1}{2}}\,,\,H^0_{I_{k,q}}+1)^{\frac{1}{2}}(S_{I_{k,q}})_{j'}
\end{equation}
are analytic in the same open disk, too, since they are obtained as products of the operators, which are analytic in $\tau$, 
\begin{equation}
(\frac{1}{H^0_{I_{k,q}}+1})^{\frac{1}{2}}(V^{(k, q-1)}_{I_{k,q}})_{j'}(\frac{1}{H^0_{I_{k,q}}+1})^{\frac{1}{2}}
\end{equation}
with the operators
\begin{equation}
\frac{1}{G_{I_{k,q}}-E_{I_{k,q}}}P^{(+)}_{I_{k,q}}(H^0_{I_{k,q}}+1)^{\frac{1}{2}}\,,\,(H^0_{I_{k,q}}+1)^{\frac{1}{2}}P^{(+)}_{I_{k,q}}\frac{1}{G_{I_{k,q}}-E_{I_{k,q}}}\,, 
\end{equation}
and the latter are analytic in $\tau$ because of the expansion in (\ref{exp-G-in}). Indeed: 
\begin{enumerate}
\item[1)]  all the summands of the series in (\ref{exp-G-in}) are analytic in the same disk by Property A  at step $(k,q-1)$; 
\item[2)]  
the series are norm convergent uniformly for $|\tau|\leq t_0$,  (see Remark \ref{comment-t0}).
\end{enumerate}
Finally, from  the proof of
Lemma \ref{unboundedlemmaA3}, the series 
\begin{equation}\label{analyticity-V}
(\frac{1}{H^0_{I_{k,q}}+1})^{\frac{1}{2}}V_{k,q}^{(k,q)}(\frac{1}{H^0_{I_{k,q}}+1})^{\frac{1}{2}}:=\sum_{j=1}^\infty \tau^{j-1}(\frac{1}{H^0_{I_{k,q}}+1})^{\frac{1}{2}}(V_{k,q}^{(k,q-1)})_j^{diag}
(\frac{1}{H^0_{I_{k,q}}+1})^{\frac{1}{2}}\end{equation}
 is then easily seen to converge uniformly in $\tau$ for  $|\tau|\leq t_0$. Thus, we can conclude that the l-h-s of (\ref{analyticity-V}) is analytic  in $\mathbb{D}_{t_0}$, too.
\\

\noindent
\emph{Case c)}

\noindent
We start by recalling that,  in this case,
\begin{eqnarray}
& &(\frac{1}{H^0_{I_{r,i}}+1})^{\frac{1}{2}}V^{(k,q)}_{I_{r,i}}(\frac{1}{H^0_{I_{r,i}}+1})^{\frac{1}{2}}\\
&:=&(\frac{1}{H^0_{I_{r,i}}+1})^{\frac{1}{2}}V^{(k,q-1)}_{I_{r, i}}(\frac{1}{H^0_{I_{r,i}}+1})^{\frac{1}{2}}\\
& &+(\frac{1}{H^0_{I_{r,i}}+1})^{\frac{1}{2}}
\Big\{\,\sum_{n=1}^{\infty}\frac{1}{n!}\,ad^{n}S_{I_{k, q}}(V^{(k,q-1)}_{I_{r,i}})\,\Big\}(\frac{1}{H^0_{I_{r,i}}+1)})^{\frac{1}{2}}\,.
\end{eqnarray}
The procedure to be used is quite similar to case b). Just as in controlling the norm in Theorem \ref{norms}, we insert the identity in the form
\begin{equation}
\charf=(H^0_{I_{r,i}}+1)^{\frac{1}{2}}(\frac{1}{H^0_{I_{r,i}}+1})^{\frac{1}{2}}\,,\,\charf =(\frac{1}{H^0_{I_{r,i}}+1})^{\frac{1}{2}}(H^0_{I_{r,i}}+1)^{\frac{1}{2}}
\end{equation}
on the left and the right side of $V^{(k,q-1)}_{I_{r,i}}$, respectively. By combining the arguments used for case b) and the estimates in  Lemma \ref{unboundedlemmaA3} leading to (\ref{bound-S}) and (\ref{S-Hest}), we derive that
\begin{equation}
S_{I_{k, q}}\quad,\quad S_{I_{k, q}}(H^0_{I_{r,i}}+1)^{\frac{1}{2}},\quad \text{and}\,\, \,(H^0_{I_{r,i}}+1)^{\frac{1}{2}}S_{I_{k, q}}
\end{equation}
are analytic in $\tau\in \mathbb{D}_{t_0}$. Hence we conclude that the series 
\begin{equation}
\sum_{n=1}^{\infty}\frac{1}{n!}\,ad^{n}S_{I_{k, q}}(V^{(k,q-1)}_{I_{r,i}})
\end{equation}
consists of analytic operators, for $\tau$  in $\mathbb{D}_{t_0}$,  which, according to the proof of  Theorem \ref{norms},  converge uniformly,  for $|\tau|\leq t_0$. Hence, $V^{(k,q)}_{I_{r,i}}$ is analytic in 
$\mathbb{D}_{t_0}$, too.
\\

\noindent
\emph{Cases d-1), d-2)}

\noindent
These two cases are very similar to case c),  and we omit the proof.

\qed

As an application of the techniques we have developed in the present work, we show that the limiting function
\begin{equation}
\lim_{N\to \infty}\frac{E_N(\tau)}{N}
\end{equation}
is well defined and analytic  in $\tau\in \mathbb{D}_{t_0}$ (as in the foregoing result),  provided the chain is invariant under translations in a sense specified below. Recall that each $\mathcal{H}_i$ is a copy of a Hilbert $\mathcal{H}$, so that, given a vector $\varphi\in \mathcal{H}$, we call $\varphi_i$ its representative in $\mathcal{H}_i$. For $j\in \mathbb{N}$, we consider the unitary operators 
\begin{equation}
\mathcal{U}_{(j)_N}\,:\,\mathcal{H}_i \to \mathcal{H}_{(i+j)_N}\quad, \quad \mathcal{U}_{(j)_N}(\varphi_i):=\varphi_{(i+j)_N}\,,
\end{equation}
where $(i+j)_N:=i+j-kN$ with $k$ the smallest number in $\mathbb{N}_0\equiv \mathbb{N}\cup \{0\}$ such that $0\leq i+j-kN \leq N$.
In particular $\mathcal{U}_{(j)_N}\Omega_i=\Omega_{(i+j)_N}$. Define
\begin{equation}
\mathcal{U}_{(j)_N}\,:\,\mathcal{H}^{(N)}\to \mathcal{H}^{(N)}
\end{equation}
as
\begin{equation}
\mathcal{U}_{(j)_N}(\mathcal{H}^{(N)}):=(\mathcal{U}_{(j)_N}\mathcal{H}_1)\otimes (\mathcal{U}_{(j)_N}\mathcal{H}_2)\otimes \dots \otimes (\mathcal{U}_{(j)_N}\mathcal{H}_{N-1})\otimes (\mathcal{U}_{(j)_N}\mathcal{H}_N)\,.
\end{equation}
\begin{prop}\label{thermo}
Under the hypotheses of Theorem \ref{main-res}, and assuming that in (\ref{complex-Hamiltonian}) the potentials $V_{I_{k,i}}$, $k\leq \bar{k}$,  depend only on the length, $k$,  of the interval $I_{k,i}$, i.e., for any $0\leq i \leq N$ and any $j$
\begin{equation}
V_{I_{k,(i+j)_N}}=\mathcal{U}_{(j)_N}V_{I_{k,i}}\mathcal{U}^*_{(j)_N}\,,
\end{equation}
the limiting function $$\varepsilon (\tau):=\lim_{N\to \infty}\frac{E_N(\tau)}{N}$$ exists for any $|\tau|\leq t_0$ and is analytic in $\tau$ in the open disc $\mathbb{D}_{t_0}$.
\end{prop}

\noindent
\emph{Proof}

Thanks to the local features of the algorithm $\alpha_{I_{k,q}}$, it is not too difficult to realise that the effective potentials $V^{(k,q)}_{I_{l,i}}$ created in the course of our construction enjoy the same covariance property assumed for the potentials $V_{I_{k,i}}$ entering the initial Hamiltonian $K_N$, i.e.,
\begin{equation}
V^{(k,q)}_{I_{k,(i+j)_N}}=\mathcal{U}_{(j)_N}V^{(k,q)}_{I_{k,i}}\mathcal{U}^*_{(j)_N}\,.
\end{equation}
Consequently,  the total energy $E_N$ admits a decomposition into a sum of the type 
$$E_N(\tau)=\sum_{l=1}^{N-1} n_l E^{(N-1,1)}_l(\tau)\,,$$
where $n_l=N-l$ is the number of subsequent sets (``intervals") of length $l$ contained in ${1,2,\ldots, N}$, and $E^{(N-1,1)}_l$ is the common expectation value of the effective potentials of length $l$ in the final step, $(N-1,1)$,  
in the vacuum vector. Notice that by construction $E^{(N-1,1)}_l(\tau)$ coincides with the energy, $E_l(\tau)$,  of the same chain but of length $l$. From Theorem \ref{analyticity} we know that this function,  $E_l(\tau)$, is analytic in  $\mathbb{D}_{t_0}$. Hence, both existence and analyticity of $\varepsilon(\tau)$ follow if the sequence of functions $\big\{\frac{E_N(\tau)}{N}\big\}_{N\in \mathbb{N}}$ is shown to be uniformly Cauchy in the disk $|\tau|<t_0$.

\noindent
To show this, note that  the estimates  on the $\|\cdot\|_{H_0}$ norms of the effective potentials readily imply that  the inequality $|E^{(N-1,1)}_l|\leq 2|\tau|^{\frac{l-1}{4}}$ holds true for any  natural number $l$. Next, let $M>N$ be two positive integers.  Then,  for any $\delta>0$ there is some $N_{\delta}$ such that,  for $N>N_{\delta}$, we can estimate
\begin{eqnarray*}
\big|\frac{E_N(\tau)}{N}-\frac{E_M(\tau)}{M}\big|&&\leq \sum_{l=1}^N \big|\frac{N-l}{N}-\frac{M-l}{M}\big||E_l| +\sum_{l=N+1}^M \frac{N-l}{M}|E^{(N-1,1)}_l|\\
&&\leq \frac{M-N}{NM}\sum_{l=1}^N \,l\cdot |E^{(N-1,1)}_l|+\sum_{l=N+1}^\infty |E^{(N-1,1)}_l|\leq \frac{2}{N}\sum_{l=1}^N l\cdot|\tau|^{\frac{l-1}{4}}+2\sum_{l=N+1}^\infty |\tau|^{\frac{l-1}{4}}\\
&&\leq 2\left( \frac{C}{N}+ R_N\right)\leq\delta\,,
\end{eqnarray*} 
 for any $\tau$ such that $|\tau|\leq t_0$, where $C$ is a universal constant and $R_N$ depends only on $N$, since
the series of functions $\sum_{l=1}^\infty |\tau|^{\frac{l-1}{4}}$ and $\sum_{l=1}^N l\cdot|\tau|^{\frac{l-1}{4}}$  converge uniformly in $|\tau|$, for $|\tau|\leq t_0$, with $t_0$ as in Theorem \ref{analyticity}.

\qed

\newpage
\setcounter{equation}{0}
\begin{appendix}
\section{Appendix}

\begin{lem}\label{formboundedness}
Assume that $\|V^{(k,q-1)}_{I_{r,i}}\|_{H_0} \leq |\tau |^{\frac{r-1}{4}}$ and define
\begin{equation}
\Delta_{I_{k,q}}:=\frac{1-8|\tau|\,\sum_{j=1}^{\infty}(j+1)\,|\tau|^{\frac{j-1}{4}}}{2}\,,
%\Delta_{I_{k,q}}:=2\inf\{|E|\,;\,E\in  \text{spec} \left (P^{(+)}_{I_{k,q+1}} (G_{I_{k,q+1}}-E_{I_{k,q+1}}) P^{(+)}_{I_{k,q+1}}  \vert_{P^{(+)}_{I_{k,q+1}}\mathcal{H}^{(N)}} \right )\}\,,
\end{equation}
then for $|\tau|$  sufficiently small and independent of $k$, $q$, and $N$,
\begin{equation}
\left\|\frac{1}{G_{I_{k,q}}-E_{I_{k,q}}}P^+_{I_{k,q}} (H^0_{I_{k,q}}+1)^{\frac{1}{2}}\right\|\leq \frac{\sqrt{2}}{\Delta_{I_{k,q}}}\,.\label{op-norm-G-2}
\end{equation}
\end{lem}

\noindent
\emph{Proof.}

\noindent
The proof  follows from the Neumann expansion in (\ref{exp-G-in})-(\ref{exp-G-fin}),  from the estimate in (\ref{est-exp-in})-(\ref{est-exp-fin}), and from the spectral theorem.
\qed

\begin{lem}\label{unboundedlemmaA3}
Assume $\|V^{(k,q-1)}_{I_{r,i}}\|_{H_0} \leq |\tau |^{\frac{r-1}{4}}$ and that $|\tau|$ is sufficiently small such that  $\Delta_{I_{k,q}}\geq \frac{1}{2}$ (see Corollary \ref{cor-gap}). Then, for arbitrary $N$, $k\geq 1$, and $q\geq 2$, the inequalities
\begin{equation}\label{bound-V}
\|V^{(k,q)}_{I_{k,q}}\|_{H_0}\leq 2\|V^{(k,q-1)}_{I_{k,q}}\|_{H_0}\,,
\end{equation}
\begin{equation}\label{bound-S}
\|S_{I_{k, q}}\|\leq A\,|\tau|\,
 \| V^{(k,q-1)}_{I_{k,q}}\|_{H_0}\,,
\end{equation}
and
\begin{equation}\label{S-Hest}
\|S_{I_{k,q}}(H^0_{I_{k,q}}+1)^{\frac{1}{2}}\|=\|(H^0_{I_{k,q}}+1)^{\frac{1}{2}}S_{I_{k,q}}\| \leq B|\tau|\,
 \| V^{(k,q-1)}_{I_{k,q}}\|_{H_0}
\end{equation}
hold true for universal constants $A$ and $B$. For $q=1$,   $ V^{(k,q-1)}_{I_{k,q}}$  is replaced by $V^{(k-1,N-k+1)}_{I_{k,q}}$ in the right side of (\ref{bound-V}), (\ref{bound-S}), and (\ref{S-Hest}).
\end{lem}

\noindent
\emph{Proof}

\noindent
In the following we assume $q\geq 2$; if $q=1$ an analogous proof holds. We recall that 
\begin{equation}
V^{(k,q)}_{I_{k,q}}:= \sum_{j=1}^{\infty}\tau^{j-1}(V^{(k,q-1)}_{I_{k,q}})^{diag}_j \,
\end{equation}
and
\begin{equation}
S_{I_{k,q}}:=\sum_{j=1}^{\infty}\tau^j(S_{I_{k,q}})_j\,,
\end{equation}
with $$(V^{(k,q-1)}_{I_{k,q}})_1=V^{(k,q-1)}_{I_{k,q}}$$ 
and, for $j\geq 2$,
\begin{eqnarray}
& &(V^{(k,q-1)}_{I_{k,q}})_j\,:=\label{formula-v_j-bis}\\
& &\sum_{p\geq 2, r_1\geq 1 \dots, r_p\geq 1\, ; \, r_1+\dots+r_p=j}\frac{1}{p!}\text{ad}\,(S_{I_{k,q}})_{r_1}\Big(\text{ad}\,(S_{I_{k,q}})_{r_2}\dots (\text{ad}\,(S_{I_{k,q}})_{r_p}(G_{I_{k,q}}) \Big)\\
& &+\sum_{p\geq 1, r_1\geq 1 \dots, r_p\geq 1\, ; \, r_1+\dots+r_p=j-1}\frac{1}{p!}\text{ad}\,(S_{I_{k,q}})_{r_1}\Big(\text{ad}\,(S_{I_{k,q}})_{r_2}\dots (\text{ad}\,(S_{I_{k,q}})_{r_p}(V^{(k,q-1)}_{I_{k,q}}) \Big)\,,\quad\quad\quad\quad\,.
\end{eqnarray}
and
\begin{eqnarray}
(S_{I_{k,q}})_j:
%&=&ad^{-1}\,G_{I_{k,q}}\,((V^{(k,q-1)}_{I_{k,q}})^{od}_j)\\
&=&\frac{1}{G_{I_{k,q}}-E_{I_{k,q}}}P^{(+)}_{I_{k,q}}\,(V^{(k,q-1)}_{I_{k,q}})_j\,P^{(-)}_{I_{k,q}}\\
& &-P^{(-)}_{I_{k,q}}\,(V^{(k,q-1)}_{I_{k,q}})_j\,P^{(+)}_{I_{k,q}}\frac{1}{G_{I_{k,q}}-E_{I_{k,q}}}\,.
\end{eqnarray}
where $j\geq 1$.

\noindent
From the lines above, we have that
\begin{eqnarray}
\text{ad}\,(S_{I_{k,q}})_{r_p}(G_{I_{k,q}})
&=&\text{ad}\,(S_{I_{k,q}})_{r_p}(G_{I_{k,q}}-E_{I_{k,q}})\nonumber \\
%&=&\,[\frac{1}{G_{I_{k,q}}-E_{I_{k,q}}}P^{(+)}_{I_{k,q}}\,(V^{(k, q-1)}_{I_{k,q}})_{r_p}\,P^{(-)}_{I_{k,q}}\,,\,G_{I_{k,q}}-E_{I_{k,q}}]+h.c.\\
&=&-P^{(+)}_{I_{k,q}}\,(V^{(k,q-1)}_{I_{k,q}})_{r_p}\,P^{(-)}_{I_{k,q}}-P^{(-)}_{I_{k,q}}\,(V^{(k, q-1)}_{I_{k,q}})_{r_p}\,P^{(+)}_{I_{k,q}}\,.
\end{eqnarray}

\noindent
We start by showing the following inequality:
\begin{equation}\label{Snorm}
\|(S_{I_{k,q}})_j\|\leq \frac{2\sqrt{2}}{\Delta_{I_{k,q}}} \| (V_{I_{k,q}}^{(k, q-1)})_j\|_{H^0}\,,
\end{equation}

\noindent
where  $ \| (V_{I_{k,q}}^{(k, q-1)})_j\|_{H_0}$ will turn out to be bounded in the next step.
Going back to estimate (\ref{Snorm}), the inequality in (\ref{Snorm})  is proven by means of the following computation:
\begin{eqnarray}
& &\|(S_{I_{k,q}})_j\| \label{S-in}\\
&\leq& 2\left\|\frac{1}{G_{I_{k,q}}-E_{I_{k,q}}}P^{(+)}_{I_{k,q}}\,(V^{(k,q-1)}_{I_{k,q}})_j\,P^{(-)}_{I_{k,q}}\right\|\\
&=&2\left\|\frac{1}{G_{I_{k,q}}-E_{I_{k,q}}}P^{(+)}_{I_{k,q}}(H_{I_{k,q}}^0+1)^{\frac{1}{2}}(H_{I_{k,q}}^0+1)^{-\frac{1}{2}}(V^{(k,q-1)}_{I_{k,q}})_j(H_{I_{k,q}}^0+1)^{-\frac{1}{2}}P^{(-)}_{I_{k,q}}\right\| \\
&\leq&2\left\|\frac{1}{G_{I_{k,q}}-E_{I_{k,q}}}P^{(+)}_{I_{k,q}}(H_{I_{k,q}}^0+1)^{\frac{1}{2}}\right\|   \|(V_{I_{k,q}}^{(k,q-1)})_j\|_{H^0}\\
&\leq & \frac{2\sqrt{2}}{\Delta_{I_{k,q}}}\|(V_{I_{k,q}}^{(k,q-1)})_j\|_{H^0} \,,\label{S-fin}
\end{eqnarray}
where we have used (\ref{op-norm-G-2}) for the last inequality.

\noindent
Analogously, making use of (\ref{op-norm-G-2}) and $(H_{I_{k,q}}^0+1)^{\frac{1}{2}}P^{(-)}_{I_{k,q}}=P^{(-)}_{I_{k,q}}$, we estimate
\begin{equation}\label{S-norm-2}
\|(S_{I_{k,q}})_{j}(H^0_{I_{k,q}}+1)^{\frac{1}{2}}\|\leq \frac{2+\sqrt{2}}{\Delta_{I_{k,q}}} \|(V_{I_{k,q}}^{(k, q-1)})_{j}\|_{H^0}\,.
\end{equation}
Next, we want to prove that
\begin{eqnarray}
\|(V^{(k,q-1)}_{I_{k,q}})_j\|_{H^0}&\leq&\label{V-ineq}\\
\sum_{p=2}^{j}\,\frac{(2c)^p}{p!}&&\sum_{ r_1\geq 1 \dots, r_p\geq 1\, ; \, r_1+\dots+r_p=j}\,\|\,(V^{(k,q-1)}_{I_{k,q}})_{r_1}\|_{H^0}\|\,(V^{(k,q-1)}_{I_{k,q}})_{r_2}\|_{H^0}\dots \|\,(V^{(k,q-1)}_{I_{k,q}})_{r_p}\|_{H^0}\nonumber \\
+2\|V^{(k,q-1)}_{I_{k,q}}\|&& \sum_{p=1}^{j-1}\,\frac{(2c)^p}{p!}\,\sum_{ r_1\geq 1 \dots, r_p\geq 1\, ; \, r_1+\dots+r_p=j-1}\,\|\,(V^{(k,q-1)}_{I_{k,q}})_{r_1}\|_{H^0}\|\,(V^{(k,q-1)}_{I_{k,q}})_{r_2}\|_{H^0}\dots \|\,(V^{(k,q-1)}_{I_{k,q}})_{r_p}\|_{H^0}\,,\nonumber 
\end{eqnarray}
where $c:= \frac{2+\sqrt{2}}{\Delta_{I_{k,q}}}(> \frac{2\sqrt{2}}{\Delta_{I_{k,q}}})$.
In order to show this,  we observe that formula (\ref{formula-v_j-bis}) giving $(V_{I_{k,q}}^{(k, q-1)})_j$ contains two sums. We first deal with the second sum, namely
$$\sum_{p\geq 1, r_1\geq 1 \dots, r_p\geq 1\, ; \, r_1+\dots+r_p=j-1}\frac{1}{p!}\text{ad}\,(S_{I_{k,q}})_{r_1}\Big(\text{ad}\,(S_{I_{k,q}})_{r_2}\dots (\text{ad}\,.(S_{I_{k,q}})_{r_p}(V^{(k,q-1)}_{I_{k,q}})\Big)\,. $$
Each summand of the above sum is in turn a sum of $2^p$ terms which, up to a sign,  are permutations of
$$(S_{I_{k,q}})_{r_1}(S_{I_{k,q}})_{r_2}\ldots (S_{I_{k,q}})_{r_p}V_{I_{k,q}}^{(k, q-1)}\,,$$
with the potential $V_{I_{k,q}}^{(k, q-1)}$ which is allowed to appear at any position.
It suffices to study only one of these products, for the others can be treated in the same way. For instance, we can 
treat $$(S_{I_{k,q}})_{r_1} V_{I_{k,q}}^{(k, q-1)}   (S_{I_{k,q}})_{r_2}\ldots (S_{I_{k,q}})_{r_p}.$$
Notice that
\begin{eqnarray}
&&\|(S_{I_{k,q}})_{r_1} V_{I_{k,q}}^{(k, q-1)}   (S_{I_{k,q}})_{r_2}\ldots (S_{I_{k,q}})_{r_p}\|_{H^0} \nonumber \\
&=&\|(H_{I_{k,q}}^0+1)^{-\frac{1}{2}} (S_{I_{k,q}})_{r_1}  (H_{I_{k,q}}^0+1)^{\frac{1}{2}}   (H_{I_{k,q}}^0+1)^{-\frac{1}{2}} V_{I_{k,q}}^{(k, q-1)}(H_{I_{k,q}}^0+1)^{-\frac{1}{2}}(H_{I_{k,q}}^0+1)^{\frac{1}{2}}    (S_{I_{k,q}})_{r_2}\ldots (S_{I_{k,q}})_{r_p}  (H_{I_{k,q}}^0+1)^{-\frac{1}{2}}\|  \nonumber \\
&\leq& \|V_{I_{k,q}}^{(k, q-1)} \|_{H^0} \|(S_{I_{k,q}})_{r_1}(H_{I_{k,q}}^0+1)^{\frac{1}{2}} \|\, \|(H_{I_{k,q}}^0+1)^{\frac{1}{2}}    (S_{I_{k,q}})_{r_2}\| \ldots \|(S_{I_{k,q}})_{r_p}\| \nonumber\\
&\leq &c^p \|V^{(k,q-1)}_{I_{k,q}}\|_{H^0} \|\,(V^{(k,q-1)}_{I_{k,q}})_{r_1}\|_{H^0}\|\,(V^{(k,q-1)}_{I_{k,q}})_{r_2}\|_{H^0}\dots \|\,(V^{(k,q-1)}_{I_{k,q}})_{r_p}\|_{H^0}\,\nonumber
\end{eqnarray}
where (\ref{Snorm}) and (\ref{S-norm-2}) have been used. 
\noindent
Putting these terms together, we get the second sum of (\ref{V-ineq}).\\

As for the first sum in (\ref{formula-v_j-bis}), i.e.,

$$\sum_{p\geq 2, r_1\geq 1 \dots, r_p\geq 1\, ; \, r_1+\dots+r_p=j}\frac{1}{p!}\text{ad}\,(S_{I_{k,q}})_{r_1}\Big(\text{ad}\,(S_{I_{k,q}})_{r_2}\dots (\text{ad}\,(S_{I_{k,q}})_{r_p}(G_{I_{k,q}}) \Big)\,,$$
we note that each of its summands is in turn the sum up to a sign of all permutations of
$$(S_{I_{k,q}})_{r_1}(S_{I_{k,q}})_{r_2}\ldots(S_{I_{k,q}})_{r_{p-1}}[-P_{I_{k,q}}^+(V_{I_{k,q}}^{(k,q-1)})_{r_p} P_{I_{k,q}}^- - P_{I_{k,q}}^-(V_{I_{k,q}}^{(k,q-1)})_{r_p}P_{I_{k,q}}^+]\,.$$ 
A very minor variation of the computation above shows that the $\|\cdot\|_{H^0}$-norm of the first sum  in  (\ref{formula-v_j-bis}) is bounded
from above by 
$$\sum_{p=2}^{j}\,\frac{(2c)^p}{p!}\sum_{ r_1\geq 1 \dots, r_p\geq 1\, ; \, r_1+\dots+r_p=j}\,\|\,(V^{(k,q-1)}_{I_{k,q}})_{r_1}\|_{H^0}\|\,(V^{(k,q-1)}_{I_{k,q}})_{r_2}\|_{H^0}\dots \|\,(V^{(k,q-1)}_{I_{k,q}})_{r_p}\|_{H^0}\,.
$$
%$S_j\doteq\|(V_{I_{k,q}}^{(k, q-1)})_j\|_{H_0}$
Henceforth,  we closely follow the proof of Theorem 3.2 in \cite{DFFR}; that is, assuming $\|V^{(k,q-1)}_{I_{k,q}}\|_{H_0}\neq 0$, we recursively define numbers $B_j$, $j\geq 1$, by the equations
\begin{eqnarray}
B_1&:= &\|V^{(k,q-1)}_{I_{k,q}}\|_{H^0}=\|(V^{(k,q-1)}_{I_{k,q}})_1\|_{H^0} \label{B1} \,,\\
B_j&:=&\frac{1}{a}\sum_{k=1}^{j-1}B_{j-k}B_k\,,\quad j\geq 2\,, \label{def-Bj}
\end{eqnarray}
with \,$a>0$\, satisfying the relation
\begin{equation}\label{a-eq}
e^{2c a}-1+ \left( \frac{e^{2c a}-2c a -1}{a}\right)-1=0
%\frac{e^{8a}-8a-1}{a}+e^{8a}-1=1\,.
\end{equation}
Using (\ref{B1}), (\ref{def-Bj}), (\ref{V-ineq}),  and an induction, it is not difficult to prove that (see Theorem 3.2 in \cite{DFFR}) for $j\geq 2$
\begin{equation}\label{bound-V-B}
\|(V^{(k,q-1)}_{I_{k,q}})_j\|_{H^0}\leq B_j\,\Big(\frac{e^{2c a}-2c a-1}{a}\Big)+2\|V^{(k,q-1)}_{I_{k,q}}\|_{H^0}\,B_{j-1}\Big(\frac{e^{2c a}-1}{a}\Big)\,.
\end{equation}
From (\ref{B1}) and (\ref{def-Bj}) it also follows that
\begin{equation}\label{bound-b}
  B_j\geq \frac{2B_{j-1}\|\,V^{(k,q-1)}_{I_{k,q}}\,\|_{H^0}}{a}\,\quad \Rightarrow\quad B_{j-1}\leq a\frac{B_j}{2\|\,V^{(k,q-1)}_{I_{k,q}}\,\|_{H^0}}\,,
\end{equation} 
which, when combined with (\ref{bound-V-B}) and (\ref{a-eq}), yields
\begin{equation}\label{bound-b-bis}
B_j\geq \|\,(V^{(k,q-1)}_{I_{k,q}})_j\|_{H^0}\,.
\end{equation} 
The numbers $B_j$ are the Taylor's coefficients of the function
\begin{equation}
f(x):=\frac{a}{2}\cdot \left(\,1-\sqrt{1- (\frac{4}{a}\cdot \|V^{(k,q-1)}_{I_{k,q}}\|_{H^0}) \,x }\,\right)\,,
\end{equation}
(see  \cite{DFFR}). We observe that
\begin{eqnarray}
\|(V^{(k,q-1)}_{I_{k,q}})^{diag}_j \|_{H^0}&=& \max\{\|P^{(+)}_{I_{k,q}}(V^{(k,q-1)}_{I_{k,q}})_j P^{(+)}_{I_{k,q}}\|_{H^0}\,,\,\|P^{(-)}_{I_{k,q}}(V^{(k,q-1)}_{I_{k,q}})_j P^{(-)}_{I_{k,q}}\|_{H^0}\}\\
& =&\max_{\#=\pm }\,\|(\frac{1}{H^0_{I_{k,q}}+1})^{\frac{1}{2}}P^{(\#)}_{I_{k,q}}(V^{(k,q-1)}_{I_{k,q}})_j P^{(\#)}_{I_{k,q}}(\frac{1}{H^0_{I_{k,q}}+1})^{\frac{1}{2}}\|\\
&=&\max_{\#=\pm }\,\|P^{(\#)}_{I_{k,q}}(\frac{1}{H^0_{I_{k,q}}+1})^{\frac{1}{2}}(V^{(k,q-1)}_{I_{k,q}})_j (\frac{1}{H^0_{I_{k,q}}+1})^{\frac{1}{2}}P^{(\#)}_{I_{k,q}}\|\\
&\leq & \|(V^{(k,q-1)}_{I_{k,q}})_j \|_{H^0}\,.
\end{eqnarray}
Therefore the radius of analyticity, $t_0$,  of 
\begin{equation}
\sum_{j=1}^{\infty}\tau^{j-1}\|(V^{(k,q-1)}_{I_{k,q}})^{diag}_j \|_{H^0}=\frac{1}{\tau}\,\Big(\sum_{j=1}^{\infty}\tau^{j}(V^{(k,q-1)}_{I_{k,q}})^{diag}_j \|_{H^0}\Big)
\end{equation}
is bounded below by the radius of analyticity of $\sum_{j=1}^{\infty}x^jB_j$, i.e.,
\begin{equation}\label{radius}
t_0\geq \frac{a}{4\|V^{(k,q-1)}_{I_{k,q}}\|_{H^0}}\geq \frac{a}{4}\,,
\end{equation}
where we have assumed $0<|\tau |<1$ and we have invoked  the assumption $\|V^{(k,q-1)}_{I_{r,i}}\|_{H^0}\leq |\tau|^{\frac{r-1}{4}}$.
Thanks to the inequality in (\ref{Snorm}),  the same bound holds true for the radius of convergence of the series $S_{I_{k,q}}:=\sum_{j=1}^{\infty}\tau^j(S_{I_{k,q}})_j\,$\,.
For $0<|\tau|<1$ and in the interval $(0,\frac{a}{8})$, using (\ref{B1}) and  (\ref{bound-b-bis}),  we can estimate
\begin{eqnarray}
\sum_{j=1}^{\infty}|\tau |^{j-1}\|(V^{(k,q-1)}_{I_{k,q}})^{diag}_j \|_{H^0}&\leq &\frac{1}{|\tau |}\sum_{j=1}^{\infty}|\tau |^jB_j\\
&=&\frac{1}{|\tau |}\cdot \frac{a}{2}\cdot \left(\,1-\sqrt{1- (\frac{4}{a}\cdot \|V^{(k,q-1)}_{I_{k,q}}\|_{H^0}) \,|\tau|}\,\right)\\\
&\leq &(1+C_a \cdot |\tau | )\,\|V^{(k,q-1)}_{I_{k,q}}\|_{H^0}\,,
\end{eqnarray}
for some $a$-dependent constant $C_a>0$.
%From (\ref{bound-V-B}) we derive that, for $j\geq 2$,
%\begin{equation}\label{bound-series}
%\|(V^{(k,q-1)}_{I_{k,q}})_j\|\leq 2\|V^{(k,q-1)}_{I_{k,q}}\|\,B_{j-1}\Big(\frac{e^{8a}-1}{a}\Big)\,.
%\end{equation}
%Using (\ref{bound-series}) and $B_1=\|\,V^{(k,q-1)}_{I_{k,q}}\|=\|\,(V^{(k,q-1)}_{I_{k,q}})_1\|$, we find that
%\begin{eqnarray}
%\|V^{(k,q)}_{I_{k,q}}\| & = & \|\sum_{j=1}^{\infty}t^{j-1}(V^{(k,q-1)}_{I_{k,q}})^{diag}_j \|\\
%&\leq &\|\,V^{(k,q-1)}_{I_{k,q}}\,\|+2\|\,V^{(k,q-1)}_{I_{k,q}}\,\|\cdot \Big(\frac{e^{8a}-1}{a}\Big)\sum_{j=2}^{\infty}|t|^{j-1}B_{j-1}\,.\\
%& &+B_j\,\Big(\frac{e^{8a}-8a-1}{a}\Big)
%\end{eqnarray}
Hence the inequality in (\ref{bound-V})  holds true, provided  $|\tau |$ is sufficiently small but independent of $N$, $k$, and $q$. 
%For, from the convergence of  $\sum_{j=1}x^jB_j$ in the interval $[0,\frac{a}{4})$, we get 
%$$\sum_{j=2}^{\infty}|t|^{j-1}B_{j-1}<\frac{1}{2\Big(\frac{e^{8a}-1}{a}\Big)}\,,$$ 
%for any $0<|t|<\bar{t}$, where $\bar{t}$ only depends on $a$.  
 In a similar way, we derive (\ref{bound-S}) and (\ref{S-Hest}), using  (\ref{S-in})-(\ref{S-fin}) and (\ref{S-norm-2}), respectively.
 \qed

\end{appendix}

\end{document}